\documentclass[preprint,aps,nofootinbib,tightenlines,amsfonts]{revtex4}
\usepackage{epsfig}
\usepackage{bm}

\def\CO{{\cal O}}
\def\OMIT#1{{}}

\def\si{^1 \hskip -0.04in S _0}
\def\siii{^3 \hskip -0.04in S _1}
\def\diii{^3 \hskip -0.04in D _1}
\def\pislash{ {\pi\hskip-0.6em /} }
\def\pislashsmall{ {\pi\hskip-0.375em /} }
\def\nopi{ {\rm EFT}(\pislash) }
\newcommand{\gsim}{\raisebox{-0.7ex}{$\stackrel{\textstyle >}{\sim}$ }}
\newcommand{\lsim}{\raisebox{-0.7ex}{$\stackrel{\textstyle <}{\sim}$ }}

\def\tr{di-baryon}

\def\lone{l_1}
\def\ltwo{l_2}
\def\loneA{l_{1,A}}

\begin{document}

\begin{figure}[!t]
\vskip -1.5cm
\leftline{{\epsfxsize=1.5in \epsfbox{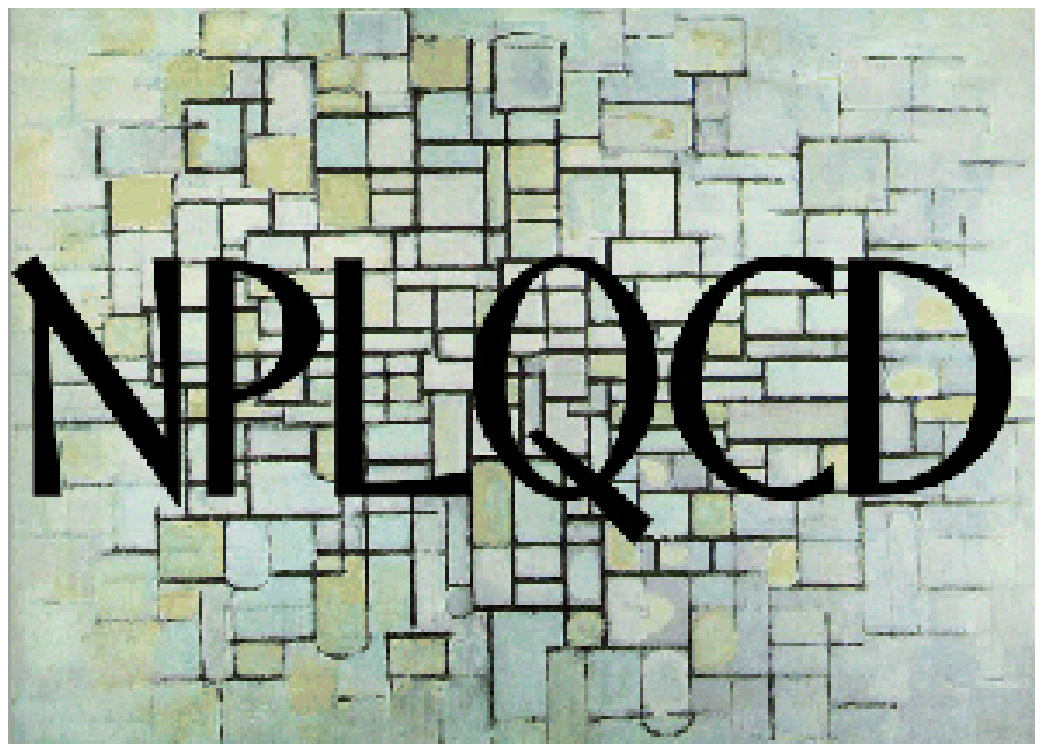}}}
\end{figure}

\preprint{\vbox{ \hbox{NT@UW-04-004} }}

\vphantom{}

\title{Electroweak Matrix Elements in the Two-Nucleon Sector from
  Lattice QCD} \author{William Detmold and Martin J.  Savage}
\affiliation{ Department of Physics, University of Washington,
  Seattle, WA 98195-1560.}

\vphantom{} \vskip 0.5cm
\begin{abstract} 
  \vskip 0.5cm
\noindent 
We demonstrate how to make rigorous predictions for electroweak matrix
elements in nuclear systems directly from QCD.  More precisely, we
show how to determine the short-distance contributions to low-momentum
transfer electroweak matrix elements in the two-nucleon sector from
lattice QCD.  In potential model descriptions of multi-nucleon
systems, this is equivalent to uniquely determining the meson-exchange
currents, while in the context of nuclear effective field theory, this
translates into determining the coefficients of local,
gauge-invariant, multi-nucleon-electroweak current operators.
The energies of the lowest-lying states of two
nucleons on a finite volume lattice with periodic boundary conditions
in the presence of a background magnetic field are sufficient to
determine the local four-nucleon operators that contribute to the
deuteron magnetic moment and to the threshold cross-section of
$np\rightarrow d\gamma$.  
Similarly, the energy-levels of two nucleons
immersed in a background isovector axial weak field 
can be used to
determine the coefficient of the 
leading local four-nucleon operator contributing to the
neutral- and charged-current break-up of the deuteron.
This is  required for the extraction of
solar neutrino fluxes at SNO and
future neutrino experiments.  
\vskip 1.0cm \leftline{\today}
\end{abstract}

\maketitle

\vfill\eject

\section{Introduction}

A major goal of nuclear physics is to be able to rigorously
compute the properties and interactions of nuclei directly from QCD.
While QCD is formulated in terms of quarks and gluons, the relevant
degrees of freedom of strongly interacting systems probed at
low-momentum are the lowest-lying hadrons.  While the properties of
these hadrons are well known experimentally, and some of their
interactions are also well studied (such as the nucleon-nucleon (NN)
cross-section), in general, direct measurement of the interactions
that contribute to nuclear processes is not possible.  For instance,
there are multi-nucleon interactions induced at the chiral symmetry
breaking scale, that will not be directly measured, but which
contribute to the interactions of nuclei.  The only way of rigorously
computing strong interaction observables and electroweak observables
involving hadrons, is with lattice QCD. 
In lattice QCD space-time is
discretized and Green functions are evaluated in Euclidean space.  
At this point in time the variety of processes that can be addressed with lattice
QCD is quite limited.  Present day computational power restricts the
sizes of lattices, the lattice spacings, and the quark masses that can
be used in simulations, making the extraction of physical observables
non-trivial.

Even if infinite computing power were to become available tomorrow 
most of the
formal framework with which to calculate most nuclear properties and
observables does not yet exist. Notable exceptions are the NN and
nucleon-hyperon (NY) scattering lengths and range
parameters~\cite{yang,Luscher:1986pf,Luscher:1990ux,Beane:2003yx,Beane:2003da}.
At the heart of the issue is the Maiani-Testa theorem~\cite{Maiani:ca}
which precludes determination of scattering amplitudes from
Euclidean-space Green functions at infinite volume away from kinematic
thresholds .  By generalizing a result from non-relativistic quantum
mechanics~\cite{yang} to quantum field theory,
L{\"u}scher~\cite{Luscher:1986pf,Luscher:1990ux} showed how to extract
the $2\rightarrow 2$ scattering amplitude from the energy-levels of
two particles in a finite-volume lattice simulation.  This technique
has been used to determine the low-energy $\pi\pi$ phase shifts
directly from QCD, e.g. Ref.~\cite{Aoki:2002ny}.  However, only one
lattice QCD calculation of the NN scattering
lengths~\cite{Fukugita:1994ve} exists, and it is quenched with
relatively large quark masses (for a review see
Ref.~\cite{Fiebig:2002kg}).

In this work we establish a framework with which to extract the
electroweak matrix elements in the two-nucleon sector.  This is only a
small step toward a framework with which to determine electroweak
matrix elements in arbitrary multi-nucleon systems, however, the
two-nucleon sector is pleasantly challenging all by itself.  Naively
this appears to require the computation of hadronic five-point
correlators, a daunting task that is beyond the reach of current
computers.  However, calculation of the energy-levels of two nucleons
in a  finite volume in the presence of appropriate background electroweak
fields allows for these matrix elements to be extracted with
little  more computing resources than are needed to calculate
the scattering parameters.  
We start by studying the energy-levels of two nucleons 
immersed  in a background magnetic field.
Clearly a calculation of the deuteron
electromagnetic properties will not be heralded as a major
accomplishment (though it would be an impressive test of lattice QCD)
as they are well-known from precise experiments, e.g.
Ref.~\cite{Tornow:2003ze}.  However, they do provide an illustration
of how the long-distance and short-distance contributions can be
isolated with lattice QCD.  This technique is then generalized to weak
background fields in order to allow for calculation of the axial
matrix element that dominates $\nu d$ 
break-up~\cite{Chen:2002pv,Butler:2002cw,Park:2002yp}, the process used by 
SNO~\cite{Ahmad:2002jz}
to determine the flavor composition of the solar neutrino
flux.  The present-day
uncertainty in this matrix element will provide a significant
uncertainty in the determination of neutrino mass differences and
mixing matrix derived from future experiments.

Determinations of the electroweak matrix elements will be
independent of the hadronic theory used to compute the infrared (IR)
properties of QCD as long as it is a complete theory consistent with
all the symmetries of QCD.  In this work we use effective field theory
(EFT), and in particular, the pionless EFT, $\nopi$ to study the very
low-momentum ($|{\bf p}|\ll m_\pi$) interaction of nucleons and
electroweak gauge
fields~\cite{Kaplan:1998tg,Kaplan:1998we,vanKolck:1998bw,Chen:1999tn}.
$\nopi$ has been developed over the last few years to
describe systems with unnaturally large scattering lengths, as are
found in the multi-nucleon sector of QCD.  Such large scattering
lengths result from a fine-tuning that nature has presented to us,
more than likely for anthropic reasons, and as a result, the analysis
in this work will not apply for arbitrary values of the light-quark
masses.

\section{Electromagnetic Observables in the Two-Nucleon Sector and
  ``Meson-Exchange Currents''}

In the traditional approach to calculating electroweak observables in
nuclear physics, one assumes that all processes are dominated by
one-nucleon interactions with the electroweak probe, and that multi-nucleon
contributions are small.  These multi-nucleon interactions result
from currents coupling to the exchange of single mesons, $\pi$'s,
$\rho$'s and so forth, the same mesons that are used to construct
nucleon-nucleon potentials~\footnote{The modern potentials that best fit all
  available data, e.g. $AV_{18}$, 
do not rely on single meson exchange to describe the
  short-distance component of the NN potential, instead use some ``best-fit''
  functional forms that minimize the $\chi^2/dof$ of the fit to NN data.
}
and are
called ``meson-exchange currents'' (MEC's)~\footnote{Of course, this
terminology assumes that all interactions can be represented by meson
exchange, which while true in the large-$N_C$ limit of QCD, is not
true in general.}.
In the context of EFT, these matrix elements receive contributions from
one-nucleon interactions with the electroweak current, 
from interactions with the mesons that have masses
below the cut-off of the theory (in the pionless theory there are no mesonic 
contributions, while in the pionful theory there are contributions from pions)
and from local, two- (or more) nucleon -- electroweak current
interactions.  These multi-nucleon interactions result from the physics 
at momentum scales above the cut-off of the EFT that has been
integrated out.  
A significant advantage of the EFT  is that no
attempt is made to force these interactions to look like they result
from the exchange of mesons -- they are what they are.  However,
before the EFT can become predictive, the coefficients of the various
operators that occur in it must be determined either from experiment or
from lattice QCD.

As the scattering lengths in both the $\si$ and $\siii$ channels 
($a_1=-23.714~{\rm fm}$ and $a_3=5.425~{\rm fm}$)
are
unnaturally large (both channels are near their IR
fixed-points~\cite{Kaplan:1998tg,Kaplan:1998we,Birse:1998dk,Birse:1998tm}),
$\nopi$ can even be used to describe electroweak processes involving
the deuteron.  $\nopi$ can be recast into a somewhat simpler theory with
which to perform computations up to a given order in the momentum
expansion by introducing dynamical \tr\ 
fields~\cite{Kaplan:1996nv,Beane:2000fi}.  As S-matrix elements
calculated in the usual formulation of $\nopi$ and with the \tr\ 
formalism are the same at any given order in the expansion
and the two frameworks
differ only in the ultra-violet (UV), both forms will give identical
results for observables in a  finite-volume.

In this work we will not reproduce the methodology of
$\nopi$~\cite{Chen:1999tn}, however, we shall discuss relevant aspects
of the \tr\ formalism~\cite{Kaplan:1996nv,Beane:2000fi}.  In terms of
nucleon and \tr\ degrees of freedom, the leading order (LO) low-energy
strong interactions for $|{\bf p}|\ll m_\pi/2$ are described by a
Lagrange density of the form
\begin{eqnarray}
{\cal L} & = & 
N^\dagger\left[\ i\partial_0 + {\nabla^2\over 2 M}\right] N
\ -\ t_j^\dagger\left[\ i\partial_0 + {\nabla^2\over 4 M}-\Delta_3\right] t^j
\ -\ s_a^\dagger\left[\ i\partial_0 + {\nabla^2\over 4 M}-\Delta_1\right] s^a
\nonumber\\
& &
- y_3 \left[\ t_j^\dagger\  N^T P_3^j N\ +\ {\rm h.c.}\right]
- y_1 \left[\ s_a^\dagger\  N^T P_1^a N\ +\ {\rm h.c.}\right]
\ \ \ ,
\label{eq:lagST}
\end{eqnarray}
where $N$ is the nucleon annihilation operator, $t^j$ is the $\siii$
\tr\ annihilation operator with spin-index $j$, and $s^a$ is the $\si$
\tr\ annihilation operator with isospin-index $a$.  Further, $P_3^j$
is the spin-isospin projector for two nucleons in the $\siii$ channel
with spin-index $j$, while $P_1^a$ is the spin-isospin projector for
two nucleons in the $\si$ channel with isospin-index $a$,
\begin{eqnarray}
P_3^j & = & {1\over\sqrt{8}}\  \tau_2 \otimes \sigma_2\sigma^j
\ \ ,\ \ 
P_1^a \ = \ {1\over\sqrt{8}}\ \tau_2\tau^a \otimes \sigma_2
\ \ \ .
\end{eqnarray}
In the Lagrange density in eq.~(\ref{eq:lagST}), a factor of
the nucleon mass $M$ has been absorbed 
into the definition of the nucleon fields and similarly a factor of $2M$
into the \tr\ fields.  The S-wave interactions are enhanced by a
factor of the expansion parameter, $1/Q$, and are treated
non-perturbatively. However the interactions that introduce mixing
with higher partial waves, e.g. $\siii-\diii$ mixing, are suppressed by at
least $Q^2$ and so only  S-wave to S-wave interactions are required 
to the order we are working~\footnote{
The analysis would become far more involved if this suppression was not present
-- see Ref.~\cite{Li:2003jn}.}.  
In order to recover the scattering
amplitudes in both S-wave channels, the constants that appear in
eq.~(\ref{eq:lagST}) are~\cite{Kaplan:1996nv}
\begin{eqnarray}
y_3^2 & = & {8\pi\over M^2 r_3}
\ \ ,\ \ 
y_1^2\ =\ {8\pi\over M^2 r_1}
\ \ ,\ \ 
\Delta_3 \ = \ {2\over M r_3} \left( {1\over a_3}-\mu\ \right)
\ \ ,\ \ 
\Delta_1 \ = \ {2\over M r_1} \left( {1\over a_1}-\mu\ \right)
\ ,
\label{eq:tdefs}
\end{eqnarray}
where $\mu$ is the renormalization scale, $a_3$ and $r_3$ are the
scattering length and effective range in the $\siii$ channel, and
$a_1$ and $r_1$ are the scattering length and effective range in the
$\si$ channel.

For the processes under consideration, the local, gauge-invariant,
electroweak interactions between two-nucleons and the gauge fields
need to be included in addition to the interactions that result from
gauging the Lagrange density in eq.~(\ref{eq:lagST}).  The additional
electromagnetic gauge-invariant operators that contribute to the
processes of interest at LO and NLO in the power-counting
are~\cite{Kaplan:1998sz,Beane:2000fi}
\begin{eqnarray}
{\cal L} & = & {e\over 2 M} N^\dagger \left[ \kappa_0 + \kappa_1\tau^3 \right] 
\sigma\cdot {\bf B} N
\ -\ { e\over M} \left( \kappa_0  -  {\tilde \ltwo\over r_3} \right)
i\epsilon_{ijk} \ t_i^\dagger t_j B_k
\nonumber\\
&&
\ +\ {e \lone\over M\sqrt{r_1 r_3}} \ \left[ t_j^\dagger s_3 B_j + {\rm h.c.}\right]
\ \ \ ,
\label{eq:lagMag}
\end{eqnarray}
where $\kappa_p=\kappa_0+\kappa_1$ and $\kappa_n=\kappa_0-\kappa_1$
are the proton and neutron magnetic moments,
$e$ is the electric charge, and ${\bf B}$ is the
magnetic field.
The magnetic moment interaction of the $\siii$ \tr\ 
is written in such a way that if $\tilde\ltwo=0$ the deuteron magnetic
moment is the sum of the neutron and proton magnetic
moments~\cite{Kaplan:1996nv,Beane:2000fi,Kaplan:1998sz,Chen:1999tn}
(at NLO)
\begin{eqnarray}
\mu_d & = & {e\over M}\left(\ \kappa_0\ +\ {\gamma_0\over 1-\gamma_0 r_3}
  \tilde\ltwo\ \right)
\ \ \ ,
\label{eq:deutMag}
\end{eqnarray}
where $\gamma_0$ is the binding momentum of the deuteron in an 
infinite volume with no background fields, and satisfies the truncated effective range
expansion,
\begin{eqnarray}
\gamma_0\ -\ {1\over a_3}\ -\ {1\over 2} r_3 \gamma_0^2 & = & 0
\ \ \ ,
\label{eq:pole}
\end{eqnarray}
where the deuteron energy is $E=-\gamma_0^2/M$.  
As spin-dependent
interactions are suppressed in the infrared, $\tilde\ltwo$ is expected
to be small and this is verified experimentally.  To make some
connection with traditional nuclear physics, the constants
$\tilde\ltwo$ and $\lone$ correspond to multi-nucleon interactions
with the electromagnetic field induced by MEC's.

The breakdown scale of $\nopi$ restricts the magnitude of electroweak
fields that can be described.  In order to have a convergent
expansion, the magnetic field must be less than $|e {\bf B}| \ll 2
m_\pi M \sim 10^5\ {\rm MeV}^2$ so that the one-nucleon magnetic
moment interaction shifts the nucleon mass by less than $m_\pi$.
Further, requiring that the nucleon magnetic polarizability shifts the
nucleon mass by less than the magnetic moment (see
eq.~(\ref{eq:mpupshift}) below) requires that
$4\pi\frac{\beta_{p,n}}{2} B_0^2 \ll \mu_{p,n} B_0$. Using the
experimental values of $\beta_p=1.2\pm0.7\pm0.3\pm0.4 \times
10^{-4}\,{\rm fm}^3$ and $\beta_n=6.5\pm2.4\pm3.0 \times 10^{-4}\,{\rm
  fm}^3$ \cite{OlmosdeLeon:zn,Lundin:2002jy} give a constraint of 
$|e {\bf B}| \ll 1.8\times 10^5~{\rm MeV}^2$, which is a somewhat weaker
constraint than from the magnetic moment interaction.  A more
naive, but still reasonable restriction is that 
$|e {\bf B}| \ll m_\pi^2\sim 2\times10^4$~MeV$^2$ arrived at by requiring nothing more
than that ``$|e {\bf B}|$ is small'', and it is this more restrictive
limit that we  adhere to.

\section{Lattice QCD and background fields} 
\label{sec:lqcdbf}

A direct calculation of multi-nucleon processes such as $n p \to d
\gamma$ is impossible on a Euclidean space lattice.  As briefly
mentioned in the introduction, the obstruction is the Maiani-Testa
theorem \cite{Maiani:ca} which precludes the calculation of S-matrix
elements away from kinematic thresholds in an infinite volume.  In
principle however, by performing lattice calculations of matrix
elements of the relevant electroweak operators at unphysical
kinematics in appropriate multi-nucleon states, e.g., $\langle d ({\bf
  p}=0) | J^\mu | n ({\bf p}=0) p ({\bf p}=0)\rangle$, the
undetermined coefficients of the low-energy EFT (modified to
accommodate the injection of energy by the lattice) can be determined,
leading to an {\it ab initio} calculation of the low-energy dependence
of the process in question.  Unfortunately, by present day standards,
calculation of such five-point correlators is prohibitively expensive
in computational terms. In order to do such a calculation, the
two-nucleon (six quark) sources and sinks and the electroweak current
operator must be tied together with quark propagators in all possible
ways and the corresponding contributions averaged over the ensemble of
gauge configurations.  Naively, this is one or two orders of magnitude
more difficult than computing the NN four-point correlator.
Clearly, a more cost-effective solution to this problem is desired

We will show that calculations of electroweak processes, such as 
$n p \to d \gamma$ and  $\bar\nu_e d \to n n e^+$, 
will be feasible in the near future.  Our
method is to some extent motivated by the earliest lattice
calculations of the magnetic moments of the proton and neutron.
Modern calculations of these quantities are performed by taking
nucleon matrix elements of the electromagnetic current operator.
However, in the early days of lattice QCD, calculations of nucleon
matrix elements were computationally prohibitive (just as
multi-nucleon matrix elements are today) and the magnetic moments were first
extracted by measuring the shift in the nucleon mass when the lattice
is immersed in a background magnetic
field~\cite{Martinelli:1982cb,Bernard:1982yu,Rubinstein:1995hc}.
Taking the magnetic field to be constant and aligned along the
$z$-direction, ${\bf B}=B_0 \hat{\bf e}_z$ the nucleon masses are
shifted, e.g.  the mass of a proton with spin aligned in the direction
of the field is
\begin{equation}
\label{eq:mpupshift}
M_{p\uparrow}(B_0)=M + \mu_p B_0 + 4\pi\frac{\beta_{p}}{2} B_0^2
+ \CO(B_0^3)    
\ \ \ . 
\end{equation} 
However, it is the energy of the nucleon that is measured on the lattice, and
when immersed in a background magnetic field the energy eigenstates of the
proton in an infinite volume are
\begin{eqnarray}
\label{eq:Epup}
E^{(n)}_{p\uparrow}(B_0) & = & 
M 
\ +\ {|e B_0 |\over M} \left( n+{1\over 2}\right)
\ +\ {p_z^2\over 2M}
\ + \ \mu_p B_0 \ + \ 4\pi\frac{\beta_{p}}{2} B_0^2
+ \CO(B_0^3)    
\ \ \ ,
\end{eqnarray} 
where $n$ is an integer denoting the Landau-level occupied by the  proton, and 
$p_z$ is the momentum of the proton parallel to the magnetic field.
For weak magnetic fields, the spin-dependent energy-splitting
is proportional to the magnetic moment of the proton,
$\mu_p$, while  
for stronger fields the magnetic polarizability, $\beta_p$
becomes significant and measurable \cite{Zhou:2002km,Burkardt:1996vb}.  To be complete
we note that by using background electric fields, electric
polarizabilities \cite{Fiebig:1988en,Christensen:2002wh} and the
electric dipole moment of the nucleon \cite{Aoki:1989rx} have been
investigated.

To apply a spatially constant background magnetic field aligned in the
$z$-direction in a lattice simulation, one takes a SU(3) 
gauge-configuration and modifies the link fields by
\begin{equation}
  U_\mu(x)\to U_\mu(x) U_\mu^{ext}(x)
\ \ \ ,
\end{equation}
with
\begin{equation}
  U_{0,3}^{ext}(x)=1 \quad\quad U_{1}^{ext}(x)=e^{+i\beta x_2}
\quad\quad U_{2}^{ext}(x)=e^{-i\beta x_1}
\ \  \ ,
\end{equation}
where $\beta$ is proportional to $e B_0 b^2$ where $b$ is the lattice
spacing~\footnote{We work in the continuum limit, $b\to0$, throughout.}. 
A vector potential with periodic boundary conditions in the transverse 
directions requires that $eB_0 A_{xy}/2\pi\in\mathbb{Z}$, where
$A_{xy}$ is the transverse area of the box. This requirement generates
large magnetic fields for present day
lattices~\cite{Martinelli:1982cb,Bernard:1982yu,Rubinstein:1995hc},
large enough to shift the nucleon mass by $\gsim m_\pi$, and thus too strong to
be described by $\nopi$.
Not requiring
periodic boundary conditions on the vector potential introduces an
exceptional plaquette that can be placed wherever it does the least
``harm'' in the single nucleon
case~\cite{Martinelli:1982cb,Bernard:1982yu,Rubinstein:1995hc}.  It
remains to be seen if this plaquette can be placed somewhere to have
minimal impact on multi-nucleon simulations or if the periodic
requirement must be enforced.

For the weak interactions, and in particular for an axial-vector
interaction, the left- and right-handed quarks need to be immersed in
different background fields.  One way to implement this is to use
domain-wall (Kaplan~\cite{Kaplan:1992bt}) fermions in the simulation and have the
background field vary only in the fifth dimension. 
A background axial
field was used in the first lattice calculation of
the axial coupling of the nucleon~\cite{Fucito:ff}.

By applying such background electroweak fields to nucleons in a finite
volume, the energy-levels become sensitive to the various one- and
two-nucleon electroweak current operators that contribute to the
deuteron properties and interactions, such as $n p \to d \gamma$,
$\bar\nu_e d \to n n e^+$ and other related electroweak processes.  It
is interesting to note that lattice calculations of the energy-levels 
relevant to the deuteron magnetic (or weak) 
moment (isoscalar operator) will be more difficult than those 
impacting isovector processes as they 
require the evaluation of disconnected diagrams (where the operator
does not connect to valence quarks) which are significantly harder to
compute, even in the single nucleon sector \cite{Gusken:1999te}.

As we will discuss later, it will be necessary to work with asymmetric
volumes of dimensions $\eta_1 L\otimes \eta_2 L\otimes
L$~\cite{Li:2003jn}.  The energy-levels of two-particles in an
asymmetric volume have been discussed recently~\cite{Li:2003jn} and we
review this analysis and extend a number of relevant aspects in
Appendix~\ref{app:one}.  It is straightforward to show that the
locations of the low-lying energy-levels of two-nucleons interacting
in an S-wave on an asymmetric lattice are given by the solutions
of~\cite{Luscher:1986pf,Luscher:1990ux,Beane:2003yx,Beane:2003da,Li:2003jn}
\begin{eqnarray}
p\cot\delta - {1\over\pi L} S(\eta_1,\eta_2;\tilde p^2) &= & 0
\ \ \ ,
\label{eq:exactee}
\end{eqnarray}
where $p\cot\delta$ uniquely describes the infinite-volume S-wave scattering
amplitude below inelastic thresholds.  
The function
$S(\eta_1,\eta_2;\tilde p^2)$ is given by
\begin{eqnarray}
S(\eta_1,\eta_2;\tilde p^2) & = & 
{1\over \eta_1\eta_2} \sum _{{\bf n}}^{\Lambda_n} {1\over |\tilde {\bf n}|^2 -\tilde p^2}
\ -\ 4\pi\Lambda_n
\ \ \ ,
\label{eq:See}
\end{eqnarray}
where $|\tilde {\bf n}|^2 = {1\over\eta_1^2} n_1^2 + {1\over\eta_2^2}
n_2^2 + n_3^2$, and the limits of the linearly-divergent,
three-dimensional sum are given by the ellipsoid $|\tilde {\bf n}|^2
\le \Lambda_n^2$.  The energy corresponding to a given value of
$\tilde p$ is
\begin{eqnarray}
E & = & {p^2\over M}\ = \ \tilde p^2\ {4\pi^2\over M L^2}
\ \ \ .
\end{eqnarray}
Given the behavior of two-nucleon energy-levels as a function of the
volume, the phase shifts (or at least the first few parameters of the
effective range expansion $p\cot\delta=-\frac{1}{a}+\frac{r}{2}p^2
+\ldots$) can be recovered.  To aid in such recovery, one can find
analytic, large volume ($L\to\infty$) expansions for the locations of
the energy-eigenstates, as done by L{\"u}scher for the case of
$\eta_{1,2}=1$~\cite{yang,Luscher:1986pf,Luscher:1990ux,Beane:2003yx,Beane:2003da}
and done by Li and Liu~\cite{Li:2003jn} for the lowest continuum level
in asymmetric volumes.  We detail this construction in
Appendix~\ref{app:one}, and give explicit expressions for the large
volume locations of the low-lying levels in the two-nucleon sector.  A
small $L$ expansion of eq.~(\ref{eq:exactee}) can also be
performed~\cite{Beane:2003da}, though we do not do this here.

It is straightforward to extend this analysis of two-nucleon 
energy-levels to include a background electroweak 
field~\footnote{Recently, 
Bedaque~\cite{Bedaque:2004kc} has considered simulating the NN system in a
background field that couples to baryon number, but which has
vanishing field-strength.
It is found that this is equivalent to performing the simulations with twisted
boundary conditions on the quark fields, and that the energy-levels of the
system are sensitive to the phase difference between the boundaries.
This appears to be a promising avenue for future investigations.
}.  
The presence of the
background fields can give rise to a number of complications such as
Landau-levels and mixing of two-nucleon states of differing total
spin and isospin. For this reason we shall develop the necessary
generalizations of the above formulas as we need them.

The Hamiltonian of a structure-less point particle of charge $e$, moving
in  a uniform, time-independent magnetic field in the $z$-direction is
\begin{eqnarray}
\hat H & = & { |\hat {\bf p}|^2\over 2 M} 
\ + \ {1\over 2} M \omega^2 (\hat x^2+\hat y^2)
\ +\ {e B_0\over 2 M} \ \hat l_z
\ \ \ ,
\end{eqnarray}
where $\omega = |{e B_0\over 2 M}|$, and $\hat l_z$ is the angular momentum
operator parallel to the magnetic field.  
In an infinite volume, the energy-eigenvalues are given by
the expression in eq.~(\ref{eq:Epup}).
However, in a finite volume the oscillator potential is bounded above, 
and for small enough lattices the potential arising from
the magnetic field can be treated as a perturbation.
Consequently, for a small magnetic field  
we can work with momentum eigenstates in the transverse direction.
Requiring that the energy-shifts due to these magnetic interactions are small
compared with the inter-level spacing of the two-nucleon momentum eigenstates in the finite
volume gives,
\begin{eqnarray}
|e B_0| & \ll & {8\sqrt{3} \pi\over L_\perp^2}
\ \ \ ,
\label{eq:llcon}
\end{eqnarray}
where $L_\perp$ is the size of the transverse dimensions.
In what follows we work with momentum eigenstates in the transverse directions,
and thus are constrained by eq.~(\ref{eq:llcon}).

\section{Deuteron Magnetic Moment : Isolating the Two-Nucleon Contribution}

It is useful to begin with the simplest case to analyze and understand
--- the deuteron magnetic moment.  To isolate the two-nucleon
contribution from the one-nucleon contribution (the magnetic moments of
the proton and neutron combined in the $S=1$ channel), the
energy-levels of two nucleons interacting in the $\siii$ channel in a
finite volume with a background magnetic field in the $z$-direction,
${\bf B}= B_0 \ \hat {\bf e}_z$, are computed.

It is obvious from eq.~(\ref{eq:mpupshift}) that the $m=\pm 1$
magnetic sub-states of the NN system will have a linear dependence on
the magnetic field for small fields.  Indeed, the effect of the
background magnetic field on the single-nucleon states is to shift
their masses from $M$ to
\begin{eqnarray}
M_{p\uparrow} & = & M - {e B_0\over 2 M} \kappa_p
\ \ ,\ \ 
M_{p\downarrow} \ = \ M + {e B_0\over 2 M} \kappa_p
\nonumber\\
M_{n\uparrow} & = & M - {e B_0\over 2 M} \kappa_n
\ \ ,\ \ 
M_{n\downarrow} \ = \ M + {e B_0\over 2 M} \kappa_n
\ \ \ ,
\label{eq:NmassB}
\end{eqnarray}
where $\kappa_{p,n}$ are the proton and neutron magnetic moments.
These mass-shifts appear as residual mass-terms in the single nucleon
propagators that contribute to the infinite-volume NN scattering
amplitude, and hence to the position of the energy-levels of
two-nucleons in a finite volume~\footnote{ Only the free-space nucleon
  mass $M$ is removed by a phase redefinition of the nucleon field in
  defining the non-relativistic theory.  The presence of interactions
  between different spin and isospin components of the nucleon field
  prevents a rescaling of each component to completely remove the
  mass-shifts in eq.~(\protect\ref{eq:NmassB}).  }.  
In the weak-$B_0$ limit contributions to the NN energy-eigenvalues
that are quadratic and higher-order in $B_0$, such as the nucleon
polarizabilities and modifications arising from the eigenstates in the
transverse directions being Landau-levels and not momentum
plane-waves, can be neglected. 

Since it is a non-relativistic system, a simple recipe can be followed
to determine the location of the low-lying energy-eigenstates of two
nucleons in a background magnetic field in a finite
volume~\cite{Beane:2003yx,Beane:2003da}.  Construct the inverse
scattering amplitude in the background magnetic field, replace the
continuum momentum-space loop integral (arising from the
bubble-diagrams) by a discrete sum over the allowed momenta in the
volume and then find the zeros of the real
part~\cite{Beane:2003yx,Beane:2003da}.  It follows directly from this
recipe that the location of the $m=\pm 1$ energy-levels  are given by the solutions to
\begin{eqnarray}
p\cot\delta_3 \ -\ 
{1\over \pi L} S(\eta_1,\eta_2; \tilde p^2 \pm \tilde u_0^2)
\mp {e B_0\over 2} \left( \tilde\ltwo - r_3 \kappa_0 \right)
& = & 0
\ \ \ ,
\label{eq:mpmone}
\end{eqnarray}
where $\kappa_0={1\over 2} (\kappa_p+\kappa_n)$ and $p\cot \delta_3(
p^2)=-\frac{1}{a_3}+\frac{r_3}{2} p^2+\ldots$ is the effective range
expansion in the $\siii$ channel.  The one-nucleon contribution to the
location of the NN energy-levels is through the term
\begin{eqnarray}
\tilde u_0^2 & = & {L^2\over 4\pi^2} \ e B_0 \kappa_0
\ \ \ ,
\end{eqnarray}
and through $\kappa_0$ in the last term in eq.~(\ref{eq:mpmone}).
The one-nucleon and two-nucleon (determined by $\tilde\ltwo$)
terms enter eq.~(\ref{eq:mpmone}) differently.
Consequently, the
short-distance and the long-distance interactions with a magnetic
field can be separated, i.e.  $\tilde\ltwo$ can be isolated from
$\kappa_0$ on the lattice.  Note that the expression in
eq.~(\ref{eq:mpmone}) reduces to the expression in
eq.~(\ref{eq:exactee}) in the limit of vanishing
magnetic field.

For symmetric lattices with $\eta_{1,2}=1$,
it follows directly from eq.~(\ref{eq:mpmone}) that for $L\gg a_3$ there
are two states that become the deuteron $m=\pm 1$ states in the
infinite-volume limit, that have energies
\begin{eqnarray}
E_{-1}^{(m=\pm 1)} & = & 
-{\gamma_0^2\over M}
\left[\  1\ +\ {12\over \gamma_0 L} {1\over 1-\gamma_0 r_3}
  e^{-\gamma_0 L}\ \right]
\ \mp\ \mu_d B_0
\ \ \ ,
\label{eq:moneBSAsym}
\end{eqnarray}
in a  finite-volume.
The deuteron magnetic moment at NLO in the EFT is given in
eq.~(\ref{eq:deutMag}), but
eq.~(\ref{eq:moneBSAsym}) is true at all orders in the expansion.  The
leading finite-volume corrections in eq.~(\ref{eq:moneBSAsym}) are the same as
those at $B_0=0$~\cite{Beane:2003da}.  Although this gives a clean
lattice measurement of the deuteron magnetic moment (itself a very
good test of the lattice approach), it does not facilitate a
separation between the one- and two-nucleon contributions.

Making an expansion of eq.~(\ref{eq:mpmone}) about the 
infinite-volume limit, $E=\mp\frac{eB_0\kappa_0}{M}$, the energies of the first
pair of continuum states in the limit where $L \gg a_3$ are
\begin{eqnarray}
E_{0}^{(m= \pm 1)} & = & 
\mp {e B_0\over M} \  \kappa_0
\ +\ {4\pi A_3 \over M L^3} \left[\ 1 \ -\  c_1 {A_3\over L}
\ +\ c_2 \left({A_3\over L}\right)^2\ +\ ...\right]
\ \ \ ,
\label{eq:moneAsym}
\end{eqnarray}
where an effective scattering length is defined to be
\begin{eqnarray}
{1\over A_3} & = & {1\over a_3} \pm { e B_0 \over 2}\  \tilde\ltwo
\label{eq:A3def}
\ \ \ ,
\end{eqnarray}
and the constants $c_{1,2}$ for $\eta_{1,2}=1$ are given in
Table~\ref{table:c1c2c1pc2pee}~\cite{Luscher:1986pf,Luscher:1990ux}.
The energies of the second pair of continuum states in the limit where
$L \gg a_3$ are
\begin{eqnarray}
E_{1}^{(m= \pm 1)} & = & 
\mp {e B_0\over M} \ \kappa_0
\ +\ 
{4\pi^2\over M L^2}
\ -\ 
{24\pi\over M L^3}{1\over p\cot\tilde\delta_0}
\left(\ 1\ +\  {2\pi c_1^\prime\over L p\cot\tilde\delta_0}
+\  {4\pi^2 c_2^\prime\over L^2 (p\cot\tilde\delta_0)^2}
\ \right)
\ \ \ ,
\label{eq:moneAsym2}
\end{eqnarray}
where
\begin{eqnarray}
p\cot\tilde\delta_0 & = & 
-{1\over A_3} + {1\over 2} r_3 {4\pi^2\over L^2}
\ \ \ ,
\end{eqnarray}
the constants $c_{1,2}^\prime$ for $\eta_{1,2}=1$ are given in
Table~\ref{table:c1c2c1pc2pee}~\cite{Luscher:1986pf,Luscher:1990ux},
and $A_3$ is defined in eq.~(\ref{eq:A3def}).

The energies in eqs.~(\ref{eq:moneAsym}) and (\ref{eq:moneAsym2})
closely resemble the well-known L{\"u}scher
formulas~\cite{Luscher:1986pf,Luscher:1990ux}, with some interesting
modifications.  The levels receive shifts proportional to $\kappa_0$
corresponding to the shifts that two non-interacting nucleons would
receive in a magnetic field in an infinite volume.  In addition,
instead of the physical scattering length entering into the
$1/L$-expansion, it is the effective scattering length, $A_3$ that
appears.  The effective scattering length, defined in
eq.~(\ref{eq:A3def}), depends upon the actual scattering length and
the short-distance contribution to the NN interaction due to the
background magnetic field.  These expressions are not entirely
unexpected.  In the continuum states the nucleons do not
``see each other'' much, and the short-distance interactions are a
perturbation (as is clear from eqs.~(\ref{eq:moneAsym}) and
(\ref{eq:moneAsym2})).  The single particle energies dominate and the
short-distance interaction, which is the sum of both the purely strong
interaction and that induced by the background field, is inserted
perturbatively.  Measuring the energies of the continuum levels over a
range of lattice volumes will allow for a clean separation between the
one-nucleon and two-nucleon contributions to the deuteron magnetic moment.

\begin{figure}[!ht]
  \centerline{{\epsfysize=2.15in \epsfbox{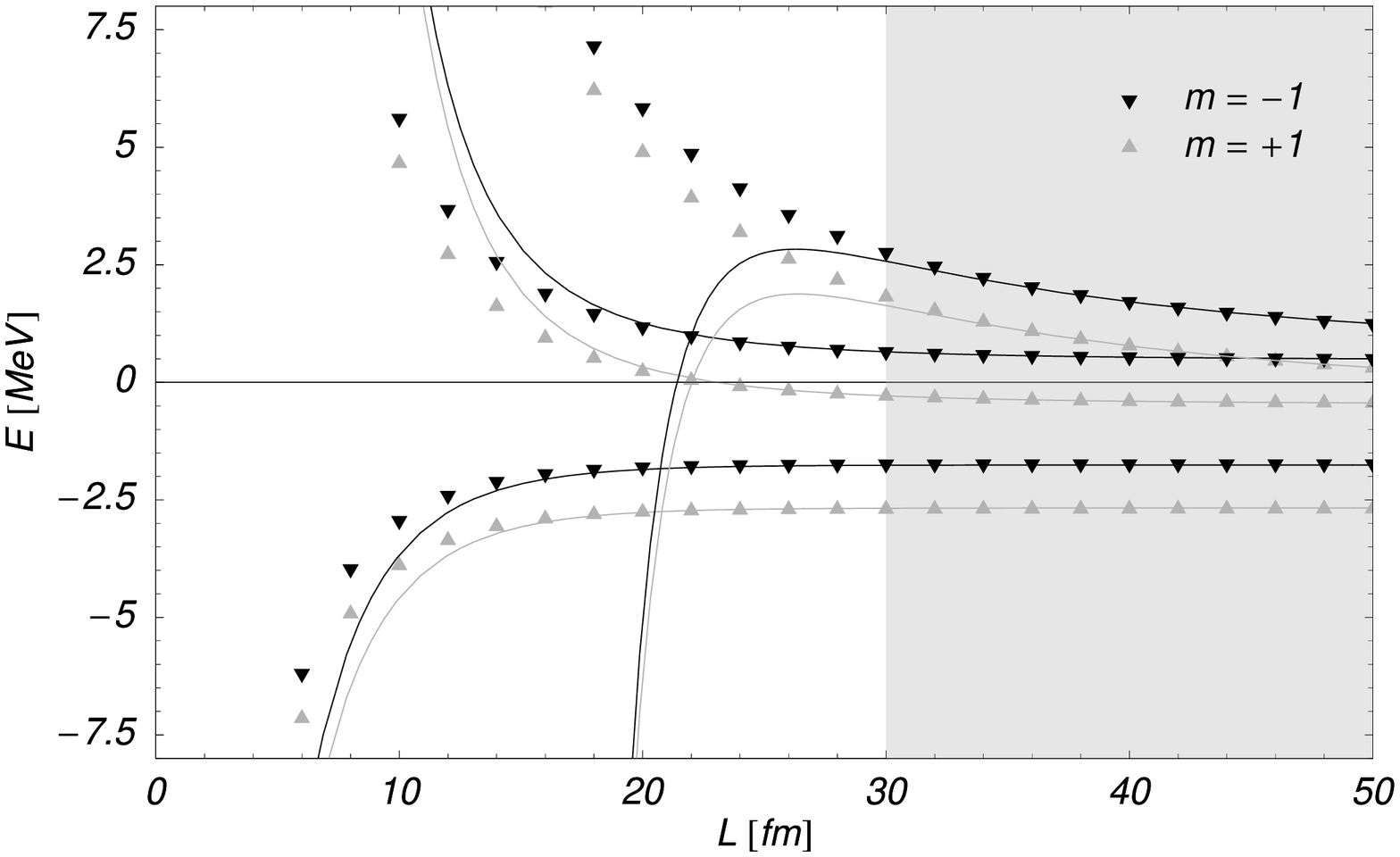}}
    {\epsfysize=2.15in \epsfbox{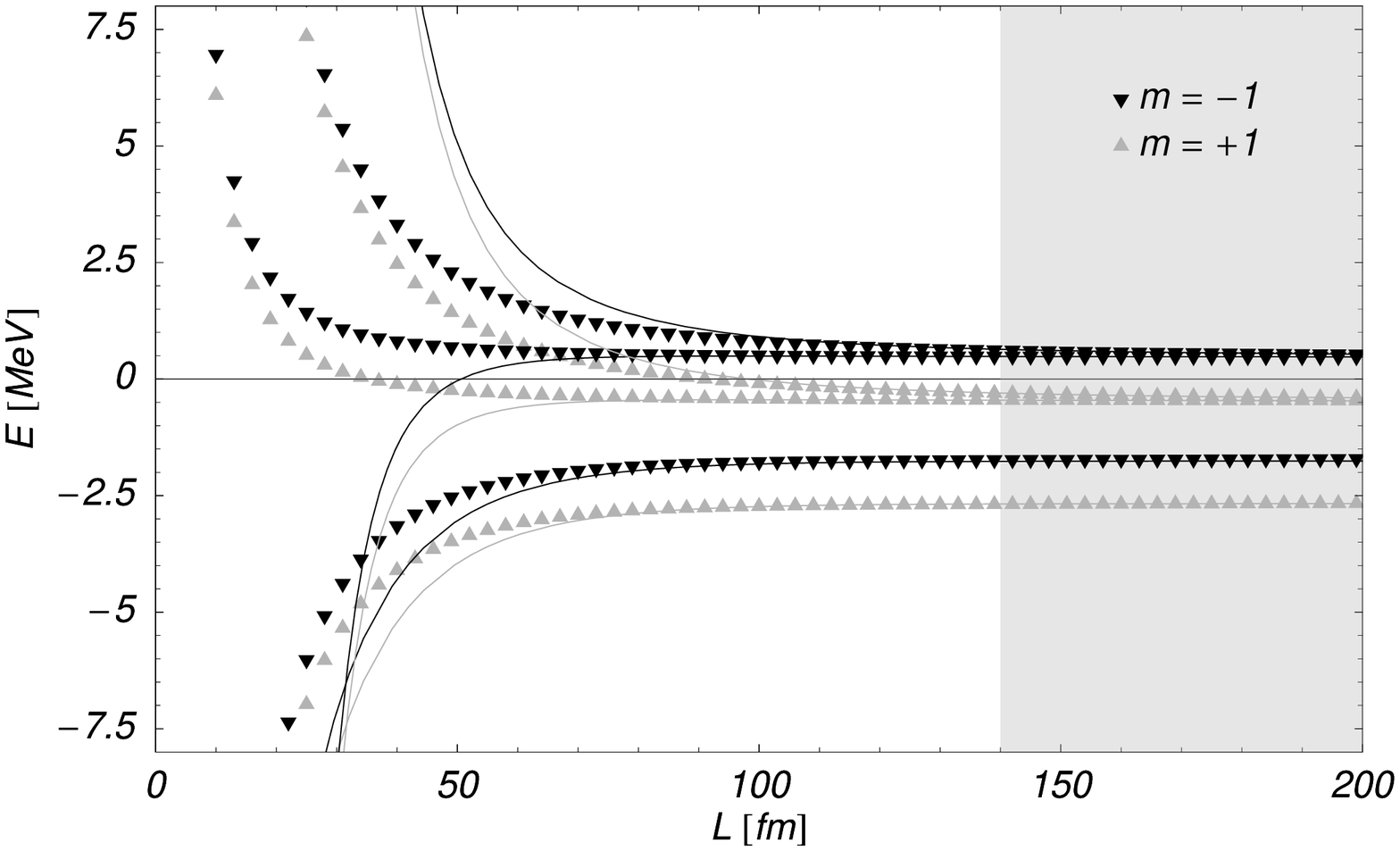}}} \vskip
  0.15in
\noindent
\caption{\it 
  The energies of the lowest-lying $\siii$ states with $m=\pm 1$ on
  the lattice as a function of the lattice size, $L$, in the presence
  of a background magnetic field of strength $eB_0=10^2~{\rm MeV}^2$.
  The left plot is for a symmetric volume, while the right plot is for
  $\eta_1=\eta_2=0.2$.  The experimental values of the nucleon and
  deuteron magnetic moments have been used.  The two types of
  triangles correspond to numerical solution of
  eq.~(\protect\ref{eq:mpmone}), and the curves correspond to the
  large-$L$ asymptotic behavior given by
  eqs.~(\protect\ref{eq:moneBSAsym}), (\protect\ref{eq:moneAsym}) and
  (\protect\ref{eq:moneAsym2}) and their asymmetric generalizations.
The gray shaded regions denote lattice sizes that are large enough so that the
effects of the transverse eigenstates being Landau-levels and not momentum
eigenstates cannot be ignored.
}
\label{fig:levelsmpmone}
\vskip .2in
\end{figure}

Given that the $\siii$ channel is very close to its non-trivial fixed
point ($a_3\to\infty$), most choices of magnetic field will move the
system away from the fixed point, and hence $|A_3|$ will typically
(i.e., for natural values of $\tilde\ltwo$, $|\tilde\ltwo|\sim1$~fm)
be dramatically smaller than $|a_3|$.  This suggests that, in
principle, lattices even smaller than those discussed in
Ref.~\cite{Beane:2003da} could be used to study the two-nucleon sector
when a background magnetic field is present.  However, nature is not
as kind as it might have been, suppressing the isoscalar
spin-dependent interactions (proportional to $\tilde\ltwo$) at
low-momentum by even more than the naive $1/N_C^{|I-J|}$ obtained in
the large-$N_C$ limit~\cite{Kaplan:1995yg}.  The experimental value of
the deuteron magnetic moment $\kappa_d = 0.85741$ is very close to the
single particle sum of $2 \kappa_0 = 0.87976$ which results in an
unnaturally small value of $\tilde \ltwo=-0.0576~{\rm fm}$ in
eq.~(\ref{eq:deutMag}).  Therefore, for magnetic fields that allow for
an analysis of the two-nucleon sector within $\nopi$, the effective
scattering length, $A_3$, is only moderately different from the actual
scattering length, $a_3$.  Specifically, for $e B_0=10^4~{\rm MeV^2}$,
$A_3$ differs from $a_3$ by $10\%$ (for the physical value of
$\tilde\ltwo$), an amount that should produce observable shifts in the
continuum two-particle energy levels calculated in lattice
simulations.  In fig.~\ref{fig:levelsmpmone} the energy-levels of the
three lowest-lying $m=\pm 1$ states are shown, along with the analytic
expressions in eqs.~(\protect\ref{eq:moneBSAsym}),
(\protect\ref{eq:moneAsym}), (\protect\ref{eq:moneAsym2})
and their asymmetric generalizations
that are valid for $L\gg a_3$.

An interesting ``twist'' could be applied to the systems we are
considering.  It is possible that the light-quark masses, $m_q$, could
be tuned to values for which the physical scattering lengths in the
$\siii$ or $\si$ channels, or both, become
infinite~\cite{Beane:2002vq,Beane:2002xf,Epelbaum:2002gb,Braaten:2003eu}.
As both channels have unnaturally large scattering lengths at the
physical values of $m_q$, only small changes in these masses will be
required, and hence the values of the operators in the pionless theory
describing the unphysical system, will differ from their physical
values by only small amount.  An advantage of considering such an
unphysical system is that $A_3$ will depend only on $\tilde\ltwo$ and
its determination on the lattice would provide a measurement of
$\tilde\ltwo$ for the physical theory with a reasonably small
systematic uncertainty.

\section{$np\rightarrow d\gamma$ : Isolating the Two-Nucleon Contribution}

The radiative capture process $np\rightarrow d\gamma$, and its inverse
reaction (photo-disintegration of the deuteron) provide a classic
demonstration of the importance of MEC's in nuclear physics.  The
one-nucleon contribution to thermal neutron capture provides only $\sim
90\%$ of the experimentally measured cross section.  From the EFT
perspective, it is natural that two-nucleon electromagnetic interactions
can contribute to this process, in the same way that they contribute
to the deuteron magnetic moment.  The cross sections for
$np\rightarrow d\gamma$ for cold neutrons and for the
photo-disintegration of the deuteron near threshold are both sensitive
to $\lone$.  
The radiative-capture cross section near threshold is given 
by~\cite{Beane:2000fi}
\begin{eqnarray}
\sigma (np\rightarrow d\gamma)
& = & 
{4\pi\alpha \left(\gamma_0^2+|{\bf p}|^2\right)^3\over M^4\gamma_0^3 |{\bf p}|}
\left[\ |\tilde X_{M1}|^2 + |\tilde X_{E1}|^2 \ +\ ...\right]
\ \ \ ,
\label{eq:npdg}
\end{eqnarray}
where contributions from multipoles other than $M1$ or $E1$, denoted
by the ellipses, are parametrically suppressed at low-energies.  Near
threshold, this process takes the two-nucleons between the $\si$
channel and the $\siii$ channel and the cross section is dominated by
the $M1$ amplitude~\cite{Chen:1999bg,Rupak:1999rk}
\begin{eqnarray}
\tilde X_{M1} & = & 
{1\over\sqrt{1-\gamma_0 r_3}}
{1\over -{1\over a_1} + {1\over 2} r_1 |{\bf p}|^2 - i |{\bf p}|}
\left[
 {\kappa_1 \gamma_0^2\over \gamma_0^2+|{\bf p}|^2}
\left( \gamma_0 - {1\over a_1} +  {1\over 2} r_1 |{\bf p}|^2\right)
+ 
\lone {\gamma_0^2\over 2}\ \right]
\ .
\label{eq:XM1amp}
\end{eqnarray}
It is worth stressing that this very simple expression for the cross
section (including the analytic form for the $E1$
amplitude~\cite{Chen:1999bg,Rupak:1999rk}) describes all available
low-energy data with high
precision~\cite{Tornow:2003ze,Chen:1999bg,Rupak:1999rk}.

A background magnetic field aligned in the $z$-direction induces a
mixing between the $\si$, $I_z=0$ state and $\siii$, $m=0$ state.
Therefore, the energy-eigenstates in the background field will be a
mixture of both the $\si$ and the $\siii$ channels, which gives rise
to quite interesting behavior.  In large volumes, the continuum
states, for which the strong interactions play only a small role, are
perturbatively close to the single particle eigenstates $p^\uparrow
n^\downarrow$ and $n^\uparrow p^\downarrow$, with energies dictated by
the single particle interactions with the background field, i.e.
linear in $B_0$ for small $B_0$.  Hence, the energies of the continuum
states will be insensitive to the value of $\lone$.  However, the
energies of the near threshold bound-states, the deuteron and the
virtual bound di-nucleon state in the $\si$ channel, for which strong
interactions play an important role, will depend quadratically on
$B_0$ for small $B_0$ and will have sensitivity to $\lone$.  This
results from the fact that these states have equal admixtures of the
single particle eigenstates in infinite volumes.  Thus, it is clear
that we will need to observe energy-shifts that are quadratic in $B_0$
in order to extract $\lone$.  This presents a potential problem that
we did not encounter for the energy shifts of the $m=\pm 1$ states.
In the directions transverse to ${\bf B}$, Landau-levels are the
single-particle eigenstates rather than momentum plane-waves.  
As discussed in section~\ref{sec:lqcdbf},
this
is a complication in infinite volumes, but becomes a far more serious
issue in finite volumes for systems with H(3) symmetry.  In order to
avoid having to contend with this problem, we eliminate it;
highly asymmetric lattices, for which $\eta_{1,2}\ll 1$, 
so as to satisfy eq.~(\ref{eq:llcon}),
strongly suppress
excitations in the transverse direction, leaving the system dominated
by fluctuations in the $z$-direction.  Ideally, one would take the
transverse dimensions almost to zero to eliminate these contributions
almost entirely.  However, the pion mass sets a lower-limit of
$\eta_{1,2} L \sim 2~{\rm fm}$ 
in order for the pionless theory to provide a
valid description.  On a symmetric lattice, as studied in
Refs.~\cite{yang,Luscher:1986pf,Luscher:1990ux,Beane:2003yx,Beane:2003da},
the lowest-lying states are in the $A_1$ representation of $H(3)$ and
as such their energy is dependent upon NN interactions with
$J=0,4,6,8,..$~\cite{Mandula:ut} with the notable omission of
interactions with $J=2$.  On an asymmetric lattice where $H(3)$ is not
a symmetry, NN interactions with even angular momentum will
contribute.  
However, in the power-counting of $\nopi$,
higher angular momentum contributions are suppressed and can be
systematically included when required.

As the background field induces mixing between the $\si$ and $\siii$
channels, the energy-eigenvalues are found by diagonalizing the
coupled channels system.  It is straightforward to show that these
eigenvalues are solutions to
\begin{eqnarray}
\left[\ p\cot\delta_1 -{S_1 + S_2\over 2\pi L}  \right]
\left[\ p\cot\delta_3 -{S_1 + S_2\over 2\pi L}  \right]
  = 
\left[\ {e B_0 \lone \over 2} + {S_1 - S_2\over 2\pi L}  \right]^2 \ \  ,
\label{eq:m0solve}
\end{eqnarray}
where
\begin{eqnarray}
S_1 & = & S(\eta_1,\eta_2 ; \tilde p^2 + \tilde u_1^2)
\ \ ,\ \ 
S_2 \ = \ S(\eta_1,\eta_2 ; \tilde p^2 - \tilde u_1^2)
\ \ ,\ \ 
\tilde u_1^2 \ = \ {L^2\over 4\pi^2} \ e B_0 \ \kappa_1
\ \ \ ,
\label{eq:s1s2}
\end{eqnarray}
$\kappa_1=(\kappa_p-\kappa_n)/2$ and $p\cot\delta_1=-\frac{1}{a_1} +
\frac{r_1}{2}p^2 + \ldots$ is the $\si$ effective range expansion. The
$\tilde u_1$ contributions result from the one-nucleon interactions with
the background field, while the two-nucleon--background 
field interactions are described by $\lone$.

\begin{figure}[!t]
  \centerline{{\epsfysize=2.15in \epsfbox{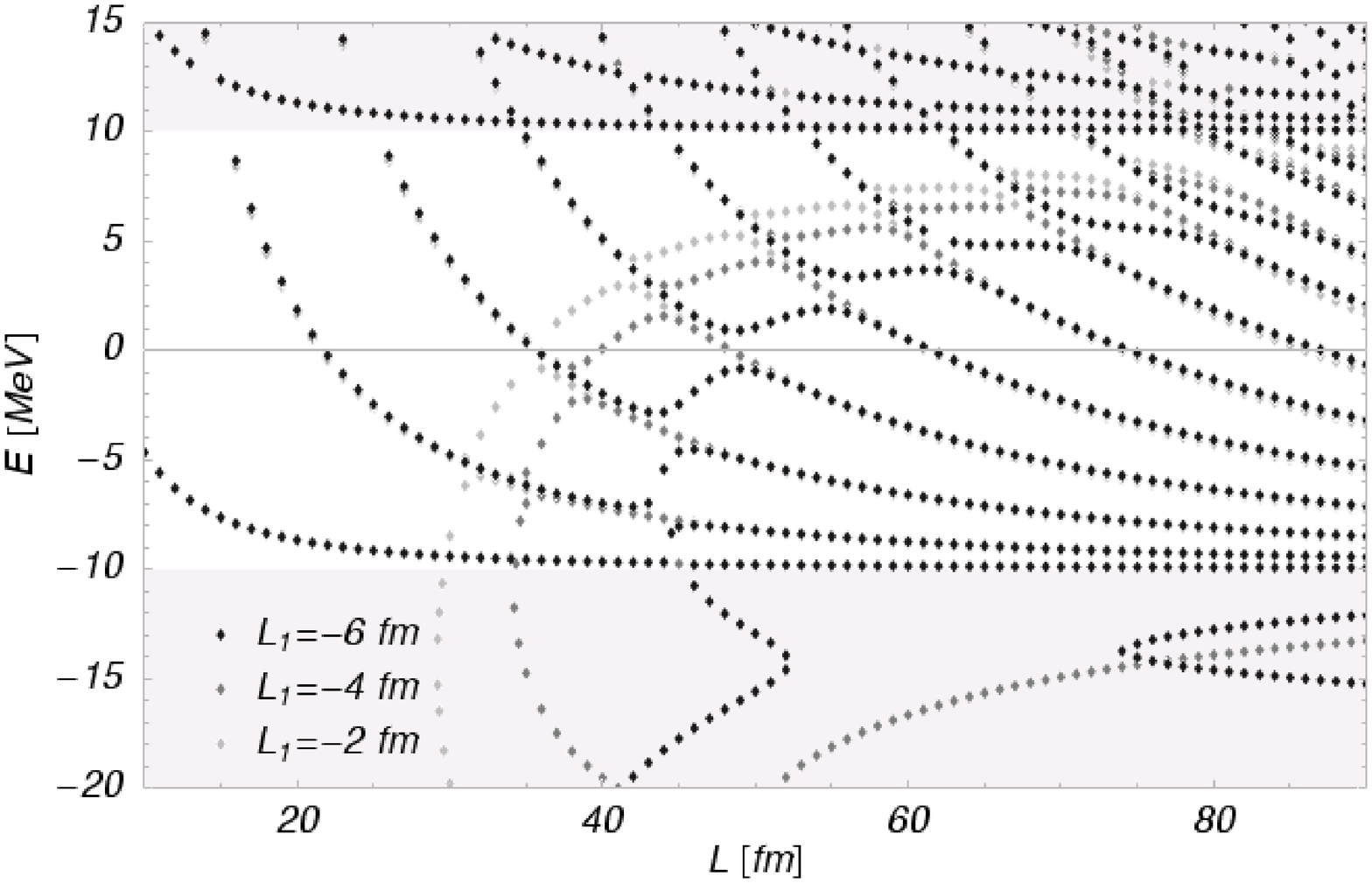}}
    {\epsfysize=2.15in \epsfbox{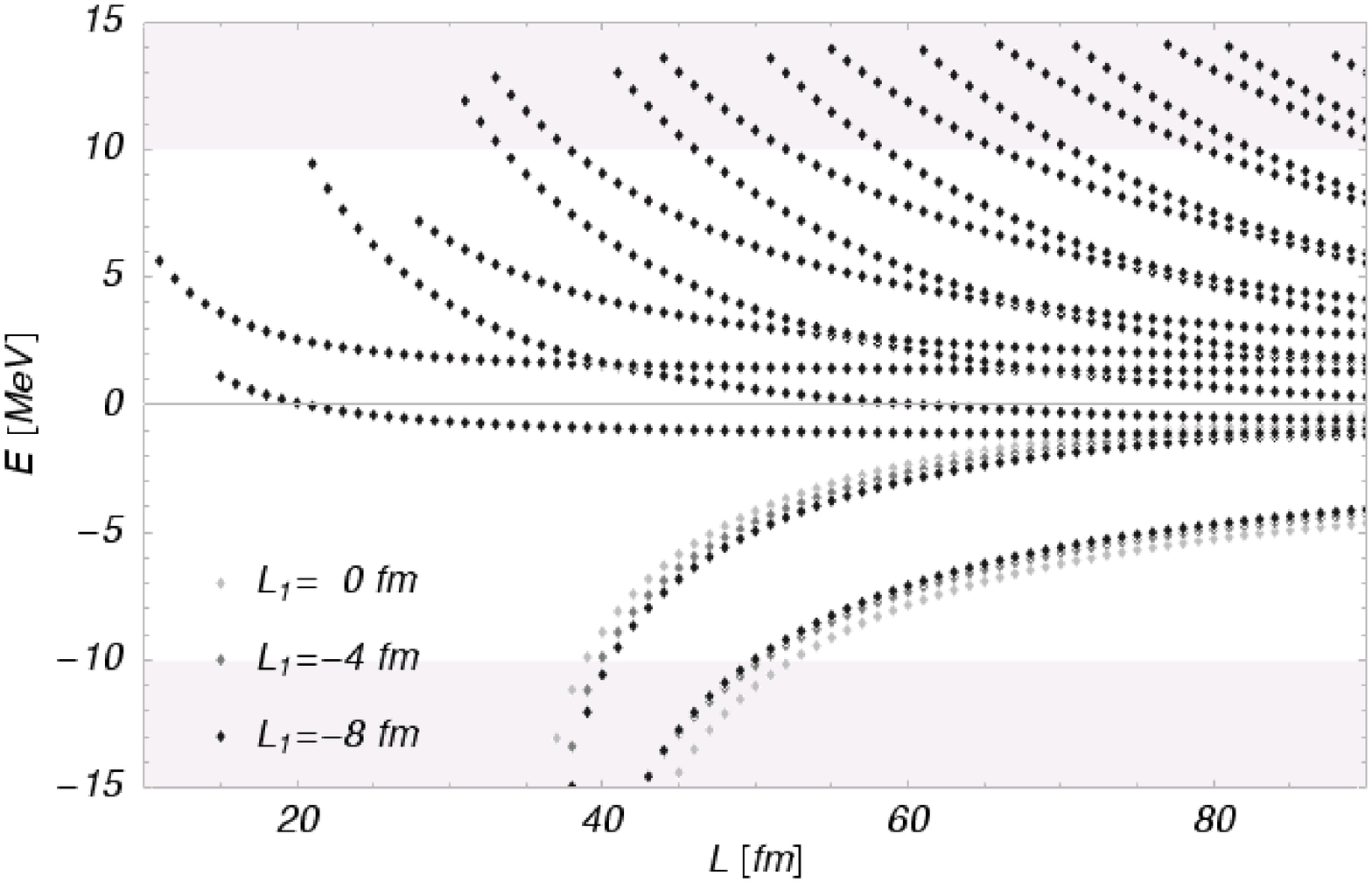}}} \vskip 0.15in
\noindent
\caption{\it 
  The energy of the lowest-lying states in the coupled 
$\si (I_z=0)-\siii (m=0)$
channels on 
an asymmetric lattice with $\eta_{1,2}=0.1$
as a function of the lattice size,
$L$, in the presence of a background magnetic field.  The left panel shows the spectrum 
for $eB_0=4\times 10^{3}~{\rm
    MeV}^2$, while the right panel shows the spectrum for
  $eB_0=5\times 10^{2}~{\rm MeV}^2$. The shaded regions indicate
  energies outside of the $\nopi$ and states therein should be treated
  with extreme caution.
}
\label{fig:levelsmzero}
\vskip .2in
\end{figure}

The $L \gg a_{1,3},r_{1,3}$ asymptotic region is more complicated in
these mixed channels than in a single channel and we do not have
practicable analytic expressions for the energies of the lowest-lying
levels in this limit.  Thus, all of our intuition and understanding
has come about through numerical solution of eq.~(\ref{eq:m0solve}).
In fig.~\ref{fig:levelsmzero}, we show the spectra of
energy-eigenvalues of the  $\si (I_z=0)-\siii (m=0)$ NN system
in magnetic fields of $eB_0=4\times 10^{3}~{\rm MeV}^2$ and 
$eB_0=5\times 10^{2}~{\rm MeV}^2$ 
on an asymmetric lattice with $\eta_{1,2}=0.1$ as a function
of lattice size, $L$.  As a function of $B_0$ and $\lone$, the
continuum levels behave precisely as one would expect, and are very
insensitive to $\lone$.  However, as anticipated, there are one or two
states in the theory (depending on the choice of parameters) which are
very sensitive to $\lone$. The location of these states in an infinite
volume are finely tuned (near the IR fixed point), and as such,
``squeezing them'' into a volume that is small in the transverse
directions destroys this fine-tuning.  However, the background field,
through the operator with coefficient $\lone$, can restore the
fine-tuning and make the location of the state depend strongly upon  $\lone$.
Further, the observation of two bound-states in the spectrum that are
sensitive to $\lone$ for some values of background field and for some
volumes indicates that the virtual bound state in the $\si$ channel
can be moved onto the first sheet by the background magnetic field.
In the infinite volume limit one of the $\lone$-dependent states
becomes the physical deuteron.  However, in a  finite volume and in the
background field its composition, and the composition of the other
states, will be quite complicated, each being a combination of
the ``deuteron'', the $\si$-``di-nucleon',' and the continuum states,
as deduced from fig.~\ref{fig:levelsmzero}.  
As $\nopi$ is applicable only at very low-energies, the energy-spectra in
fig.~\ref{fig:levelsmzero} are reliable only 
for $| E | \lsim 10~{\rm MeV}$.

For $eB_0=4\times 10^{3}~{\rm MeV}^2$, as shown in the left panel of
fig.~\ref{fig:levelsmzero}, there is one ``reliable'' state~\footnote{By
  reliable we mean that it lies within the range of validity of $\nopi$.} 
that moves to
higher energy as a function of increasing volume, and is very
sensitive to $\lone$ in the physical range of $\lone$.  For lattice
sizes between $\sim 30~{\rm fm}$ and $\sim 50~{\rm fm}$ 
in the longitudinal direction,
this level is
one of the low-lying ones in the spectrum, and therefore, somewhat
easier to extract from lattice calculations, than if it were high in
the spectrum.  Also present in this spectrum is the curious feature of
``level pair-production'' near the cut-off of the theory.  In the
absence of a background field there are two solutions to
eq.~(\ref{eq:pole}), one corresponding to the deuteron and one to an
unphysical state lying outside the theory, nonetheless both solutions
exist, and the one outside the theory is simply discarded as
unphysical.  The same line of reasoning applies to the deep states in
fig.~\ref{fig:levelsmzero}.  We interpret these states as the unbound
di-nucleon state in the $\si$ channel mixing  with continuum
states and the deuteron through the background magnetic field,
bringing an additional state down from the cut-off into the spectrum.
Its sensitivity to $\lone$ suggests that it is a strongly correlated
NN pair, as opposed to predominantly continuum in nature.
For $eB_0=5\times 10^{2}~{\rm MeV}^2$, as shown in the
right panel of fig.~\ref{fig:levelsmzero}, the two lowest-lying states
in the spectrum are both sensitive to $\lone$.  It might well be the
case that a precise measurement of the lowest-lying state in the spectrum
at weaker fields is the easiest way to determine $\lone$ from lattice
QCD.

A general statement can be made about the NN system in background
magnetic field.  There are values of the magnetic field, lattice
asymmetry and lattice size for which the spectrum of
energy-eigenvalues allows for the short-distance and long-distance
contributions to electromagnetic matrix elements to be separated, as
is clear from fig.~\ref{fig:levelsmzero}.  While a lattice extraction
of these quantities will be very important to perform and provide a
proof in principle of this method, it will not, unfortunately, tell us
anything we didn't already know experimentally.

\section{Weak Interaction Observables in the Two-Nucleon Sector}

Of central importance to present day and future neutrino
experiments is the ability to separate the flavors of neutrinos
entering the detectors.  The method to accomplish this implemented by
the SNO collaboration is to use the weak-disintegration of the
deuteron, e.g. $\nu_x d\rightarrow \nu_x np$ ($x=e,\mu,\tau$) and
$\overline{\nu}_e d\rightarrow n n e^+$.  These processes ``look''
like the photo-disintegration of the deuteron discussed in the
previous section when the electromagnetic field is replaced by a weak
gauge field.  There are both one-nucleon and two-nucleon contributions to
these processes, but unlike the electromagnetic case, the two-nucleon
contributions have significant  uncertainties associated (for a nice
discussion see Ref.~\cite{Chen:2002pv}) with them that will impact the
future determinations of the neutrino mass and mixing matrices.
Therefore a determination of the short-distance contributions to the
matrix elements for weak-disintegration of the deuteron from lattice
QCD is important.

The procedure for determining the two-nucleon weak interactions is
similar to that described in the electromagnetic sector with a few
minor differences.  For the electromagnetic interactions it sufficed
to use a background magnetic field.  However, for the weak
interactions it is convenient to introduce a fictitious
$SU(2)_L\otimes SU(2)_R$ weak gauge symmetry under which the up- and
down-quarks transform, giving rise to the interactions
\begin{eqnarray}
{\cal L}^{\rm int.} & = & 
-{1\over 4} \left[\ 
g_L W^3_{\mu,L}\left( 
\overline{u}\gamma^\mu (1-\gamma_5) u - 
\overline{d}\gamma^\mu (1-\gamma_5) d \right)
\right.\nonumber\\
&&\left.
\qquad\ +\ 
g_R W^3_{\mu,R}\left( 
\overline{u}\gamma^\mu (1+\gamma_5) u - 
\overline{d}\gamma^\mu (1+\gamma_5) d \right)
\ \right]
\ \ \ ,
\end{eqnarray}
and give vevs to $W^3_{z,L}$ and $W^3_{z,R}$, such that $\langle g_L
W^3_{z,L}\rangle = - \langle g_R W^3_{z,R}\rangle = gW$.  
Therefore, in
this background weak field there is an additional contribution to the
strong interaction Lagrange density of
\begin{eqnarray}
\delta {\cal L} & = & 
-{1\over 2} gW 
\left( 
\overline{u}\gamma^z\gamma_5 u - \overline{d}\gamma^z\gamma_5 d
\right)
\ \ \ .
\label{eq:wkbkgd}
\end{eqnarray}
This interaction leads to
interactions in the \tr\ formulation of $\nopi$ of the form
\begin{eqnarray}
\delta {\cal L} & = & 
- gW \ {g_A\over 2}\  N^\dagger\sigma^z \tau^3 N
\ -\ {gW\  \loneA  \over 2M\sqrt{r_1 r_3}} \ \left[ t_3^\dagger s_3 + {\rm
    h.c.}\right]
\ +\ ...
\ \ \ ,
\label{eq:wkeft}
\end{eqnarray}
where the ellipses denotes terms higher order in $\nopi$, and
$g_A=1.26$ is the nucleon axial coupling constant.  
We note that
due to the nature of the interaction, weak ``Landau-levels'' 
are not present in  this system.

The cross sections
for the weak-disintegration of the deuteron and other two-nucleon weak
processes depend upon the dimensionless coefficient $\loneA$.  These
processes have been studied extensively by Butler and
Chen~\cite{Butler:1999sv,Butler:2001jj} and also by Butler, Chen and
Kong~\cite{Butler:2000zp} and our (renormalization-group invariant)
$\loneA$ is related to the constants they introduce.  The coefficient
of the four-nucleon operator in $\nopi$, $^\pislashsmall L_{1,A}$, is
related to $\loneA$ via
\begin{eqnarray}
\loneA & = & 
-{2 (\mu-\gamma)\over  C_0^{(\si)}(\mu) }
\left[ \ ^\pislashsmall L_{1,A}(\mu) - \pi g_A \left( {M\over 2\pi} C_2^{(\si)}(\mu) 
+ {r_3\over (\mu-\gamma)^2}\right)\ \right]
\ \ \ ,
\label{eq:l1arel}
\end{eqnarray}
where $C_{0,2}^{(\si)}(\mu)$ are the coefficients of the zero- and
two-derivative strong interaction operators in $\nopi$.  
Numerically, eq.~(\ref{eq:l1arel})  reduces to
\begin{eqnarray}
\loneA & = & -13.4 + 0.27\  ^\pislashsmall L_{1,A}
\ \ \ ,
\end{eqnarray}
when the
RG-scale $\mu=m_\pi$ is chosen, and 
where $ ^\pislashsmall L_{1,A}$ is in units of ${\rm fm}^3$.  
Calculations have been done in
which a chiral expansion of the weak currents are performed in a
manner consistent with Weinberg's power-counting~\cite{Weinberg:rz}.
Matrix elements of these operators are taken between wave-functions
generated with the best modern NN potentials~\cite{Park:2002yp}.  This
method is somewhat {\it ad hoc}, but it has been demonstrated to be
convergent where it has been tested.  A calculation of tritium
$\beta$-decay in such a framework leads to ~\cite{Park:2002yp}
$^\pislashsmall L_{1,A}(m_\pi)=+4.2\pm 0.1~{\rm fm}^3$.  Chen, Heeger
and Robertson~\cite{Chen:2002pv} have surveyed the experimental and
theoretical constraints on $^\pislashsmall L_{1,A}(m_\pi)$ and have
also provided a model-independent determination of this quantity from
the SNO and Super-Kamiokande data, finding $^\pislashsmall
L_{1,A}=+4.0\pm6.3~{\rm fm}^3$.

The cross sections for the processes $\nu d\rightarrow \nu np$,
$\overline{\nu} d\rightarrow \overline{\nu} np$, and $\overline{\nu}
d\rightarrow e^+ nn$ are all given in terms of $^\pislashsmall L_{1,A}
(m_\pi)$, 
in Refs.~\cite{Butler:1999sv,Butler:2000zp,Butler:2001jj},
and analytic expressions for each can be found 
there.
As these
expressions are quite complicated, we do not reproduce them here.
Numerical values of the cross sections for these processes can be
found in Refs.~\cite{Butler:1999sv,Butler:2000zp,Butler:2001jj}.
As an example,
at $E_{\nu,\overline{\nu}}=10~{\rm MeV}$ the cross sections are, in
units of $10^{-42}~{\rm cm}^2$,
\begin{eqnarray}
\sigma (\nu_x d\rightarrow \nu_x np) & = & 1.76\  +\ 0.056\ \loneA
\ \ , \ \ 
\sigma (\overline{\nu}_x d\rightarrow \overline{\nu}_x np) \ = \ 1.66\
+\ 0.052\ \loneA
\nonumber\\
\sigma (\nu_e d\rightarrow e^- pp) & = & 4.07\ +\ 0.12\ \loneA
\ \ ,\ \ 
\sigma (\overline{\nu}_e d\rightarrow e^+ nn) \ = \ 1.93\ + \ 0.059\ \loneA
\ \ \ \ ,
\end{eqnarray}
where $x=e,\mu,\tau$.  Thus, a determination of $\loneA$ at the
$\sim 10\%$ level translates into an uncertainty in these cross
sections at the few percent level.

\begin{figure}[!t]
  \centerline{{\epsfysize=2.15in \epsfbox{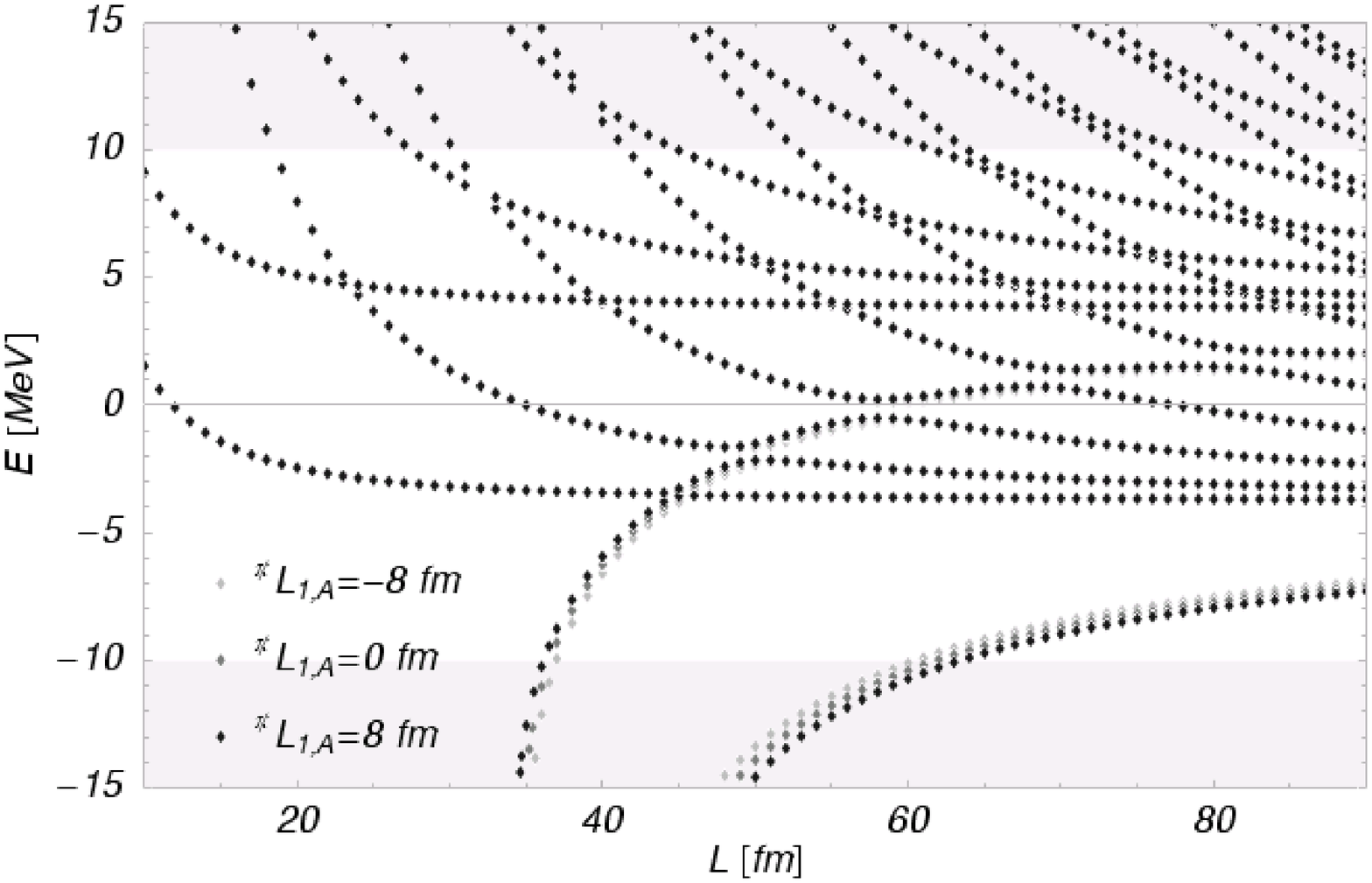}}
    {\epsfysize=2.15in \epsfbox{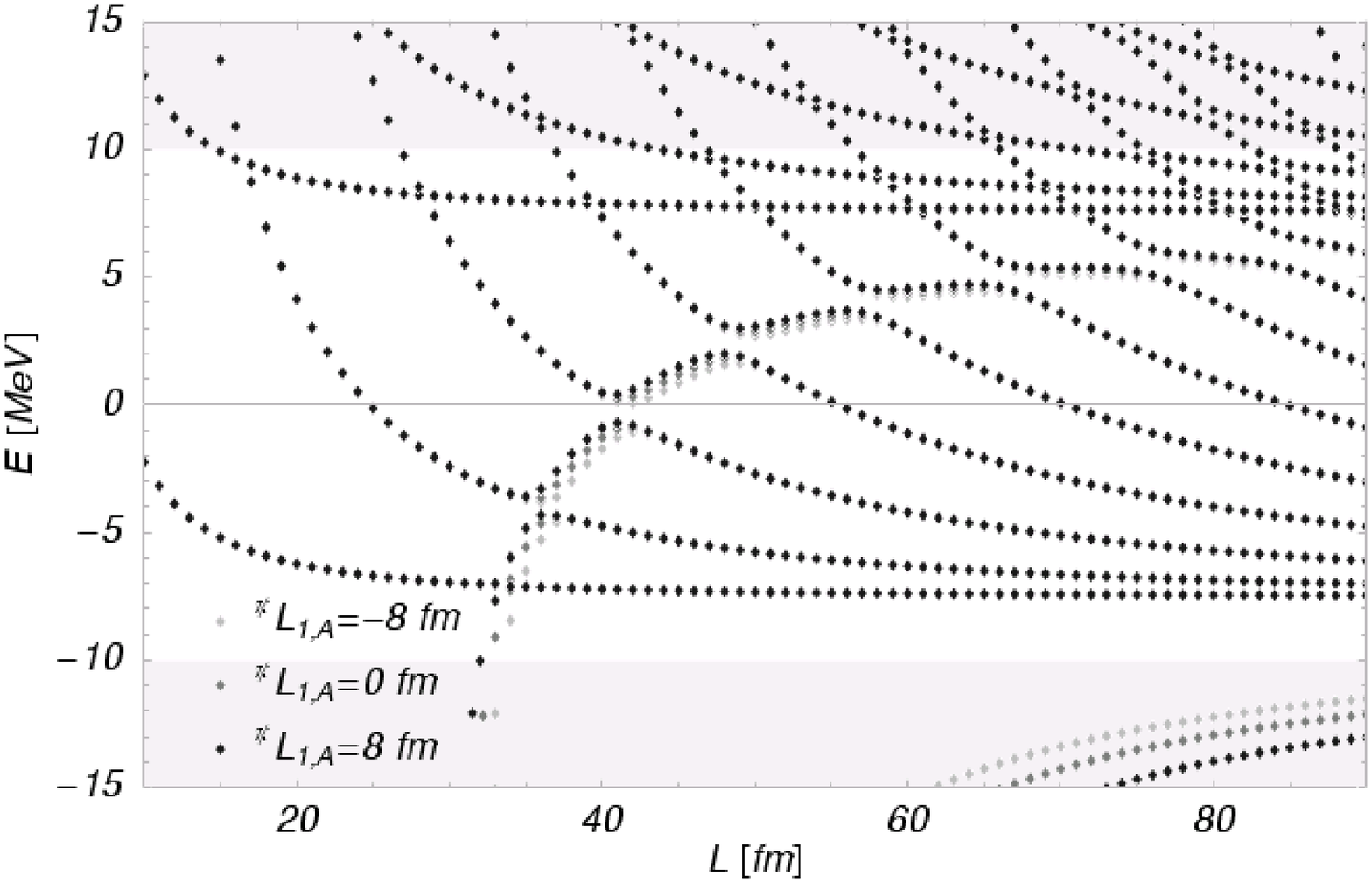}}} \vskip 0.15in
\noindent
\caption{\it 
  The energy of the lowest-lying states in the coupled 
$\si (I_z=0)-\siii (m=0)$ channels on an asymmetric lattice with $\eta_{1,2}=0.1$ as a
  function of the lattice size, $L$, in the presence of a background
  weak field.  The left panel shows the spectrum in the presence of a
  background field $gW=3~{\rm MeV}$, while the right panel shows the
  spectrum for $gW=6~{\rm MeV}$. The shaded regions indicate
  energies outside of the $\nopi$ and states therein should be treated
  with extreme caution. }
\label{fig:levelsmzerowk}
\vskip .2in
\end{figure}

In the presence of a background weak field of the form given in
eqs.~(\ref{eq:wkbkgd}) and (\ref{eq:wkeft}), the $I_z=0$ component of
the $\si$ channel mixes with the $m=0$ component of the $\siii$
channel, as is the case in the presence of a background magnetic
field.  The energy-eigenvalues of $\si (I_z=0)-\siii (m=0)$ NN system
in a finite
volume with this weak field are solutions to
\begin{eqnarray}
\left[\ p\cot\delta_1 -{S_1 + S_2\over 2\pi L} \right]
\left[\ p\cot\delta_3 -{S_1 + S_2\over 2\pi L}  \right]
\  = \
\left[ {gW\  \loneA \over 4} - {S_1 - S_2\over 2\pi L}  \right]^2
\ \ \ ,
\label{eq:m0solvewk}
\end{eqnarray}
where
\begin{eqnarray}
S_1 & = & S(\eta_1,\eta_2 ; \tilde p^2 + \tilde w_1^2)
\ \ ,\ \ 
S_2 \ = \ S(\eta_1,\eta_2 ; \tilde p^2 - \tilde w_1^2)
\ \ ,\ \ 
\tilde w_1^2 \ = \ - {L^2\over 4\pi^2} \ gW\ g_A\ M 
\ \ \ .
\label{eq:s1s2wk}
\end{eqnarray}
In analogy with the electromagnetic case, the $\tilde w_1$
contributions result from the one-nucleon interactions with the
background weak field, while the two-nucleon interactions are described
by $\loneA$.  As in the electromagnetic case, we have explored
the spectrum of this system through numerical solution of
eq.~(\ref{eq:s1s2wk}).  The spectra of two-nucleons in background weak
fields of strengths $gW=3~{\rm MeV}$ and $gW=6~{\rm MeV}$ are shown in
fig.~\ref{fig:levelsmzerowk} as a function of $L$.  In complete
analogy with the electromagnetic case, the continuum levels are
insensitive to $\loneA$, while there is one level, or 
two levels for some parameters,
that are sensitive to $\loneA$ due to the fine-tuning
in the NN sector.  For lattices with size $L\sim 40~{\rm fm}$ and
$\eta_{1,2}=0.1$ the state that is sensitive to $\loneA$ is either the
lowest state or one of the lowest states, depending upon the choice of
field.  
Over a reasonable range of $\loneA$, the energies of the 
levels that are
sensitive to $\loneA$ vary by $\sim 1~{\rm MeV}$, which should be
measurable.

It is also worth pointing out that an asymmetric lattice is not
necessary to study these weak interactions.  It is the existence of
Landau-levels in the case of a magnetic field that forced us to
consider an asymmetric lattice to suppress excitations in the
transverse directions.  However, for the weak background field in
eqs.~(\ref{eq:wkbkgd}) and (\ref{eq:wkeft}) 
the transverse eigenstates are momentum eigenstates,
and as such a symmetric lattice is sufficient.

It is clear from fig.~\ref{fig:levelsmzerowk} that such lattice
simulations of two-nucleons in a background weak field will allow for
a determination of $\loneA$.  We find this to be quite remarkable.
As a final comment on the weak properties of the two-nucleon sector,
one could explore the weak-moment of the deuteron.  This too will have
contributions from the weak-moments of the nucleons, and also from
short-distance interactions with the neutral-current gauge-field.  By
examining the two-nucleon spectrum in the presence of an isoscalar
weak field, the short-distance interaction could be isolated from the
long-distance contribution and the deuteron weak moment could be
determined.  This too would be a spectacular accomplishment for
lattice QCD.

\section{Conclusions}

In this work we have shown that lattice simulations of the two-nucleon
sector in a  finite volume and in the presence of background electroweak
fields allow for the isolation of the short-distance, multi-nucleon
electroweak currents, traditionally called ``meson-exchange
currents''.  In lattice QCD simulations one has many more ``knobs'' to
turn to get at the physics one is interested in.  In particular, the
dimensions of the lattice can be chosen to disturb the fine-tuning in
the two-nucleon sector in a controlled way to maximize sensitivity to
the short-distance electroweak interactions.  Remarkably, there can be
one or two levels in the spectrum that are highly sensitive to these
short-distance interactions, while the continuum states 
show little sensitivity.
Consequently, for some background fields and lattice dimensions, the
short-distance interactions can be isolated from the long-distance
interactions.

One would like to
see a higher-order analysis of these finite volume observables in
$\nopi$ including the effects of the Landau-levels, one- and 
two-nucleon  polarizabilities and other relevant contributions.
Further, our analysis has been performed exclusively with the pionless EFT,
$\nopi$.  While the result are rigorous where applicable, this work
needs to be extended to a larger range of energies and momenta, which
requires solving the pionful theory in a  finite volume.  
Progress toward this goal is being made in the zero- and
single nucleon sectors, but so far little progress has
been made in the multi-nucleon sectors.  
Recently, Lee, Borasoy and Schaefer~\cite{LBS} have performed the first 
lattice simulations of the pionful EFT using Weinberg's power-counting,
with an eye to a complete numerical solution of multi-nucleon 
systems~\footnote{The formal problems associated with the chiral expansion
in Weinberg's power-counting uncovered in
Ref.~\cite{Kaplan:1996xu} (see also Ref.~\cite{Beane:2001bc})
have not been resolved in Ref.~\cite{LBS}.}.
Further, encouraging progress has been
made toward setting-up lattice simulations in 
$\nopi$~\cite{Muller:1999cp,Chen:2003vy}.

The numerical analysis we have presented in this work has been
for the physical values of the strong interaction parameters, and
consequently, for the physical values of the quark masses.  Lattice
simulations of the immediate future will not be performed at the
physical quark masses, be they quenched,
partially-quenched\footnote{Generically, the non-unitary nature of the
  quenched and partially-quenched versions of QCD is problematic when
  considering two-particle scattering. However, for energies below
  ghost meson production threshold, the partially-quenched
  theory is unitary and can be analyzed rigorously~\cite{Beane:2002np}
.} or unquenched.
It is possible that the scattering lengths in these unphysical
simulations will not be unnaturally large, and the pionless
theory will not be applicable.  Therefore, it is more than likely that
the pionful theory will need to be solved to 
more precisely
understand how the scattering parameters depend upon the light quark
masses, and to directly extract the interactions that result
from physics at the cut-off scale of the pionful theory.

This work provides the first steps toward extracting the
electromagnetic and weak properties of nuclei from lattice QCD
simulations.  We encourage lattice practitioners to explore this new
area and to determine the resources required to perform such
calculations.  As mentioned earlier, even if infinite computing power
were presently available, one knows very little about what to compute in lattice
QCD in order to make predictions about nuclear processes. Therefore, we
encourage nuclear theorists to extend this work and the work of
Refs.~\cite{yang,Luscher:1986pf,Luscher:1990ux,Beane:2003yx,Beane:2003da}
so that a rigorous computational framework for nuclear physics is in
place when the computing power becomes available.

\bigskip\bigskip

\acknowledgments

We would like to thank Silas Beane, Paulo F. Bedaque, David B.
Kaplan, David Lin, Assumpta Parreno, Steve Sharpe and Matt Wingate for
numerous discussions.  Further, we thank the other members of the NPLQCD
effort for useful input.

\appendix
\section{Two Nucleons on an Asymmetric Lattice}
\label{app:one}

The energy-levels and formalism associated with two interacting
particles on an  asymmetric lattice of 
finite volume, have recently been
studied~\cite{Li:2003jn}.  We use the asymmetric lattice to suppress
nucleon momentum modes in the directions transverse to the background
magnetic field in order to extract the electromagnetic counterterm,
$\tilde\ltwo$, i.e. to minimize the contribution from the $N^\dagger {\bf
  A}\cdot {\bm\nabla} N$ 
and $N^\dagger {\bf A}^2 N$ 
interactions which generate Landau-levels in
an infinite volume. 

Explicit evaluation of the function $S(\eta_1,\eta_2;\tilde p^2)$
defined in eqs.~(\ref{eq:exactee}) and (\ref{eq:See}) can be 
accomplished in a variety of ways.
Extending  the Chowla-Selberg
relation~\cite{Elizalde:1997jv} to asymmetric lattices, it is
easy to show that for $x\agt0$
\begin{eqnarray}
S(\eta_1,\eta_2;-x^2) & = & - 2\pi^2 x
\ +\ \pi \sum_{{\bf m}\ne {\bf 0}} {1\over |\overline{\bf m}|} e^{-2\pi |\overline{\bf m}| x}
\ \ \ ,
\label{eq:sumee}
\end{eqnarray}
where $|\overline{\bf m}|^2 = \eta_1^2 m_1^2 + \eta_2^2 m_2^2 +
m_3^2$.   In order to facilitate evaluation of
$S(\eta_1,\eta_2;\tilde p^2)$, and also to determine the
coefficients in the expansion of the energy-levels as a function of
lattice size, three sums are evaluated for a given lattice asymmetry.  
The most convergent sum 
($\sim 1/\Lambda_n^3$ in the UV) is
\begin{eqnarray}
K_{\eta_1\eta_2} & = & 
{1\over \eta_1\eta_2} \sum_{{\bf n}\ne {\bf 0}} {1\over |\tilde {\bf n}|^6}
\ \ .
\label{eq:Kdefee}
\end{eqnarray}
A second sum is
\begin{eqnarray}
J_{\eta_1\eta_2} & = & 
{1\over \eta_1\eta_2} \sum_{{\bf n}\ne {\bf 0}} {1\over |\tilde {\bf n}|^4}
\nonumber\\
& = & 
{\pi^2\over x}\left[\ 1 \ +\  
\sum_{{\bf m}\ne {\bf 0}} e^{-2\pi |\overline{\bf m}| x}\ \right]
\ -\ {1\over \eta_1\eta_2 x^4}
\ +\ 
{1\over \eta_1\eta_2}  \sum_{{\bf n}\ne {\bf 0}}
{2 x^2 |\tilde {\bf n}|^2 + x^4\over 
|\tilde {\bf n}|^4 \left( |\tilde {\bf n}|^2 + x^2\right)^2}
\ \ ,
\label{eq:Jdefee}
\end{eqnarray}
and it is important to note that $J_{\eta_1\eta_2}$ is independent of
the value of $x$, and the expression in eq.~(\ref{eq:Jdefee}) 
holds for $x\agt0$, and also converges as $\sim 1/\Lambda_n^3$.  
The third sum is
\begin{eqnarray}
I_{\eta_1\eta_2} & = & 
\lim_{\Lambda_n\to\infty} \left[{1\over \eta_1\eta_2} \sum_{{\bf n}\ne {\bf 0}}^{\Lambda_n} {1\over |\tilde {\bf n}|^2}
\ -\ 4\pi\Lambda_n\right]
\nonumber\\
& = & 
- 2\pi^2 x
\ +\ \pi \sum_{{\bf m}\ne {\bf 0}} {1\over |\overline{\bf m}|} e^{-2\pi |\overline{\bf m}| x}
\ -\ {1\over \eta_1\eta_2 x^2}
\ +\ x^2 J_{\eta_1\eta_2}
\ -\ x^4 K_{\eta_1\eta_2}
\nonumber\\
& & 
\ +\ {x^6\over \eta_1\eta_2} \sum_{{\bf n}\ne {\bf 0}} 
{1\over |\tilde {\bf n}|^6 \left( |\tilde {\bf n}|^2+x^2\right)}
\ \ \ ,
\label{eq:Idefee}
\end{eqnarray}
which is also independent of the value of $x$, and the expression in 
eq.~(\ref{eq:Idefee}) holds for $x\agt0$.
The numerical values of these sums are given in Table~\ref{table:IJKee}
for different lattice asymmetries $\eta_{1,2}$.
\begin{table}[h]
\caption{The constants $I_{\eta_1\eta_2}$, $J_{\eta_1\eta_2}$,
  $K_{\eta_1\eta_2}$ for different lattice asymmetries $\eta_1 , \eta_2$.}
\vskip 0.1in
\begin{tabular}{||c||c|c|c||}
\hline
\ \ \ $\eta_1 , \eta_2$\  \ \   &\  \ \  $I_{\eta_1\eta_2}$\ \  \   & \ \ \ 
$J_{\eta_1\eta_2}$\ \  \     
& \ \  \ $K_{\eta_1\eta_2}$ \   \ \    \\
\hline
\hline
$0.1$ , $0.1$ & 206.456 &217.884 & 203.475  \\
\hline
$0.2$ , $0.2$ & 20.9815 & 56.9542 &  50.9151  \\
\hline
$1$ , $1$ & -8.91368 & 16.5323 & 8.40191  \\
\hline
\end{tabular}
\label{table:IJKee}
\end{table}
In terms of these $\tilde p$-independent constants,
the function $S(\eta_1,\eta_2;\tilde p^2)$ can be written as
\begin{eqnarray}
S(\eta_1,\eta_2;\tilde p^2) & = & 
I_{\eta_1\eta_2}\ -\ {1\over\eta_1\eta_2\ \tilde p^2}
\ +\  \tilde p^2\  J_{\eta_1\eta_2}
\ +\ {\tilde p^4 \over \eta_1\eta_2} \sum_{{\bf n}\ne {\bf 0}}^{\Lambda_n}
{1\over |\tilde {\bf n}|^4 \left( |\tilde {\bf n}|^2-\tilde p^2\right)}
\ \ \ ,
\label{eq:Seeeasy}
\end{eqnarray}
which is valid for all lattices defined by $\eta_{1,2}$, and for all 
$\tilde p^2$.
Accuracy in the 5th or 6th significant digit can be obtained
by evaluating a few finite sums that require insignificant amounts of
computer time.
An additional check on our numerics is provided by the integral
representation
\begin{eqnarray}
S(\eta_1,\eta_2;\tilde p^2) & = & 
2 \pi^{3/2}\  e^{\tilde p^2}\  \left( 2 \tilde p^2 - 1 \right)
\ +\ 
{e^{\tilde p^2}\over\eta_1\eta_2} \sum_{{\bf n}} {  e^{-|\tilde {\bf n}|^2
  }\over |\tilde {\bf n}|^2 - {\tilde p^2}}
\nonumber\\
& & 
\ -\  
\pi^{3/2}\ \int_0^1 dt\ {e^{t \tilde p^2 } \over t^{3/2}}
\left( 4 \tilde p^4 t^2\ -\ \sum_{{\bf m}\ne {\bf 0}} e^{-\pi^2 |\overline{\bf
      m}|^2\over t} \right)
\ \ \ .
\label{eq:Sintform}
\end{eqnarray}

In the limit that all lattice dimensions are much larger than any
strong interaction length scale, including the scattering length,
approximate formula for the energy-levels can be
found~\cite{yang,Luscher:1986pf,Luscher:1990ux}.  An expression for
the energy of the first continuum state of two particles on an
asymmetric lattice has been given in this large-volume limit in
Ref.~\cite{Li:2003jn}, and we add to this work by giving expressions
for the bound state (the deuteron) and for the second continuum level.
We find that the bound state energy of two nucleons in the $\siii$
channel is
\begin{eqnarray}
E_{-1}^{(\siii)} & = & 
{\gamma_0^2\over M} \left[\ 
1 + {4\over\gamma_0 L (1-\gamma_0 r_3)}\left( e^{-\gamma_0 L} + {1\over\eta_1}
  e^{-\gamma_0 \eta_1 L}
+ {1\over\eta_2} e^{-\gamma_0 \eta_2 L}\right)
\right]
\ \ \ .
\label{eq:deutee}
\end{eqnarray}
In the relevant case of $\eta_{1,2} \ll 1$, the first 
term in the round brackets can be neglected and the second two
dominate the expression (for other cases such as $\eta_{1,2}\gg1$ the
modifications are obvious).  Physically, the dominance of the
transverse dimensions is reasonable 
for $\eta_{1,2}\ll 1$
as the contributions to the
kinetic energy of the nucleons in the deuteron from these two
directions are being reduced, moving the deuteron further from its
infrared fixed-point.  As expected for an attractive interaction, the
deuteron becomes more deeply bound as the lattice volume is reduced.

The energy of the lowest-lying continuum state, corresponding to the
lowest momentum mode in all three directions, is
\begin{eqnarray}
E_0 & = & {4\pi a\over \eta_1\eta_2 M L^3}\left[\ 
1 - c_1(\eta_1,\eta_2) \left({a\over L}\right) +
c_2(\eta_1,\eta_2)\left({a\over L}\right)^2
\ +\ \ldots\right]
\ \ \ ,
\label{eq:c1ee}
\end{eqnarray}
valid in both the $\si$ and $\siii$ channels.  The values of the
coefficients $c_1$ and $c_2$ for various values of $\eta_{1,2}$ are
given in Table~\ref{table:c1c2c1pc2pee}.
\begin{table}[h]
\caption{Values of the constants $c_{1,2}(\eta_1,\eta_2)$ and  $c_{1,2}^\prime(\eta_1,\eta_2)$
for different lattice asymmetries $\eta_1 , \eta_2$. Additional values of
$c_{1,2}(\eta_1,\eta_2)$ are given in Ref.~\protect\cite{Li:2003jn}.}
\vskip 0.1in
\begin{tabular}{||c||c|c|c|c|c||}
\hline
\ \ \ $\eta_1 , \eta_2$\  \ \   
&\  \ \  $c_1 (\eta_1 , \eta_2)$\ \  \   
& \ \ \  $c_2 (\eta_1 , \eta_2)$\ \  \     
& \ \  \  $c_1^\prime (\eta_1 , \eta_2)$ \   \ \    
& \ \  \  $c_2^\prime (\eta_1 , \eta_2)$ \   \ \    
& \ \  \  $d$ \   \ \    
\\
\hline
\hline
$0.1$ , $0.1$ & 65.7171 & 2111.12 & -3.60224 & -52.944 & 1\\
\hline
$0.2$ , $0.2$ & 6.67861 & -99.6627 & -2.3242 & 0.950737  & 1\\
\hline
$0.5$ , $0.5$ & -3.61168 & 6.65945 & 0.73400 & 0.24642  & 1\\
\hline
$1$ , $1$ & -2.837297 & 6.375183 & -0.061367 & -0.354156  & 3 \\
\hline
\end{tabular}
\label{table:c1c2c1pc2pee}
\end{table}

The next continuum level corresponds to non-zero momentum in the
longest direction(s), e.g. ${\bf p}=(0,0,\pm{2\pi\over L})$ for
$\eta_{1,2}\leq 1$.
For large $L$, the energy of this level  for $\eta_{1,2}\leq 1$ is
\begin{eqnarray}
E_1 & = & {4\pi^2\over  M L^2}
- d\ {4  \tan\delta_0\over\eta_1\eta_2 M L^2}
\left[\ 
1 + c_1^\prime(\eta_1,\eta_2) \tan\delta_0
+
c_2^\prime (\eta_1,\eta_2)\tan\delta_0
\ +\ ..\right]
\ \ \ ,
\label{eq:c2ee}
\end{eqnarray}
where the phase shift is evaluated at $|{\bf p}|^2=4\pi^2/L^2$, and
$d$ is a degeneracy factor, given in Table~\ref{table:c1c2c1pc2pee}.
Again for other choices of $\eta_{1,2}$, the corresponding expressions are
easily derived.
Values of the coefficients $c_{1,2}^\prime $ are given in
Table~\ref{table:c1c2c1pc2pee} for some choices of $\eta_{1,2}$.  The
large values of some of the $c_i$'s in Table~\ref{table:c1c2c1pc2pee}
indicate that it is the ``short'' direction(s) that are setting the
scale of the finite-volume corrections, and not the ``long''
direction(s). 

\begin{figure}[!t]
  \centerline{{\epsfysize=2.5in \epsfbox{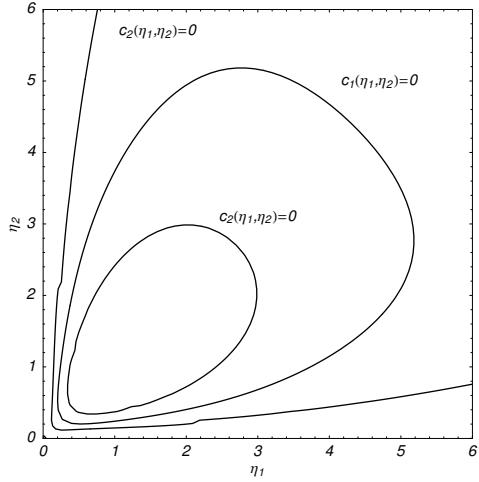}}}
\noindent
\caption{\it 
  Curves in the $\eta_1\eta_2$ plane for 
which $c_{1,2}(\eta_1,\eta_2)$ vanish. }
\label{fig:c12zeros}
\vskip .2in
\end{figure}

Finally we note that there are curves in the
$\eta_1$--$\eta_2$ plane for which $c_{1,2}(\eta_1,\eta_2)$
separately vanish, as shown in fig.~\ref{fig:c12zeros}. 
The point on this plot that is 
most relevant to our analysis is $\eta_1=\eta_2=0.268494$ for which
$c_1=0$. 
Such ``magic boxes'' may prove to be useful in lattice calculations
as the asymptotic formulas in eqs.~(\ref{eq:deutee})---(\ref{eq:c2ee}) 
have better convergence properties at smaller $L$
than those associated with arbitrary lattice volumes.
Analogous contours exist for the $c_{1,2}^\prime(\eta_1,\eta_2)$
in the $\eta_1$--$\eta_2$ plane.

\end{document}